\documentclass[a4paper,11pt]{article}

\makeatletter
\let\thermal@article@makecaption\@makecaption
\makeatother
\usepackage{jheppub}

\usepackage[T1]{fontenc} 
\usepackage{dsfont}
\usepackage{framed}
\usepackage{mathrsfs}
\usepackage{makecell}

\makeatletter
\let\@makecaption\thermal@article@makecaption
\makeatother
\usepackage{caption}
\usepackage{subcaption}
\usepackage{float}
\usepackage{needspace}

\usepackage{graphicx}
\usepackage{tikz}
\usetikzlibrary{arrows.meta,calc,decorations.pathmorphing}

\allowdisplaybreaks[1]

\let\originaltableofcontents\tableofcontents
\renewcommand{\tableofcontents}{%
  \begingroup\small
  \originaltableofcontents
  \endgroup}

\def\({\left(}
\def\){\right)}
\def\[{\left[}
\def\]{\right]}
\def\<{\langle}
\def\>{\rangle}

\def\nn{\nonumber\\}

\def\bf{\mathbf}
\def\rm{\mathrm}
\def\bb{\mathbb}
\def\cal{\mathcal}

\def\a{\alpha}
\def\b{\beta}

\def\D{\Delta}

\def\x{\xi}

\def\p{\pi}

\def\f{\phi}

\def\t{\tau}

\def\ps{\psi}

\def\Disc{\text{Disc}}
\def\Res{\text{Res}}
\def\zb{\bar z}
\def\mth{m_{\text{th}}}

\begin{document}

\title{Thermal Polyakov bootstrap}

\author[a]{Yongwei Guo,}
\author[a,b]{Zhijin Li}
\author[a]{and Tinghong Shen}
\emailAdd{guoyw33@seu.edu.cn,  zhijin\_li@seu.edu.cn, thshen@seu.edu.cn}

\affiliation[a]{School of Physics and Shing-Tung Yau Center, Southeast University,
Nanjing, 210096, China}
\affiliation[b]{Laboratoire de Physique de l'\'Ecole normale sup\'erieure,
ENS, Universit\'e PSL, CNRS, Sorbonne Universit\'e, Universit\'e Paris Cit\'e,
24 rue Lhomond, 75005 Paris, France}

\abstract{We derive thermal Polyakov bootstrap equations for CFT two-point
functions on $S^1_\beta\times\mathbb R^{d-1}$.  Combining the thermal OPE
with a dispersion relation and the method of images gives a representation
in blocks that satisfy Kubo--Martin--Schwinger (KMS) covariance.
Their local expansions contain additional thermal blocks at classical
double-twist dimensions.  Matching the prescribed physical OPE requires
cancellation of these generated contributions, yielding explicit linear
equations for thermal OPE coefficients,
including any compensating terms required by the reconstruction.
We formulate the equations for bosonic
scalar correlators and antiperiodic scalar structures extracted from
fermion correlators, and find consistent solutions in analytically tractable
thermal CFTs.  The coefficient matrix, which we call the Polyakov matrix, connects the large-spin expansion to
KMS consistency and reduces to minus the identity on the classical
double-twist spectrum.  This structure suggests a numerical approach to the
three-dimensional Ising CFT that combines known low-lying vacuum data with
a self-consistent large-spin tail.}

\maketitle 
\clearpage

\section{Introduction}
Conformal field theories (CFTs) describe fixed points of renormalization-group flows and
provide universal descriptions of classical and quantum critical phenomena.  Their
finite-temperature physics is important whenever a critical system is probed away from
the vacuum, while thermal states in holographic CFTs are dual to black-hole and
black-brane geometries.  The thermal operator product expansion (OPE) relates short-distance response functions to
vacuum CFT data and thermal expectation values, providing a common language for
field-theory, numerical, and holographic calculations
\cite{CaronHuot:2009thermal,Katz:2014rla,WitczakKrempa:2015dynamics}.  In Euclidean
signature, a CFT at inverse temperature $\beta$ lives on
$S^1_\beta\times\bb R^{d-1}$.  The thermal scale breaks part of the conformal group and
allows primary operators to acquire nonzero one-point functions; the corresponding
broken conformal Ward identities were analyzed systematically in
Ref.~\cite{Marchetto:2023xap}.  For a symmetric-traceless primary $\mathcal O$ of
dimension $\D$ and spin $J$,
\begin{align}
    \langle \mathcal O^{\mu_1\cdots\mu_J}\rangle_\beta
    =\frac{b_{\mathcal O}}{\beta^\D}
    \left(e^{\mu_1}\cdots e^{\mu_J}-\text{traces}\right),
    \label{thermal one point tensor}
\end{align}
where $e^\mu$ is the unit vector that points along the thermal circle.  Consequently, the short-distance
expansion of a thermal two-point function contains, in addition to the vacuum spectrum
and OPE coefficients, the thermal one-point coefficients $b_{\mathcal O}$.  These
coefficients constitute unknown dynamical data and are generally difficult to determine in
an interacting theory.  We denote by $a_{\mathcal O}$ the corresponding thermal OPE
coefficient, which combines $b_{\mathcal O}$ with the vacuum OPE coefficient and the
normalization of $\mathcal O$.

The Kubo--Martin--Schwinger (KMS) relation supplies the basic consistency condition on
these data \cite{Kubo:1957,MartinSchwinger:1959}.  For the correlator
$g(z,\zb)=\langle\f(z,\zb)\f(0,0)\rangle_\beta$ of two identical bosonic scalar operators $\f$
of dimension $\D_\f$, translation and spatial-rotation symmetry allow it to be
written, after setting $\beta=1$, as
\begin{align}
    g(z,\zb)=g(1-z,1-\zb),\qquad
    z=\t+i|\bf x|,\quad \zb=\t-i|\bf x|.
    \label{intro thermal KMS}
\end{align}
Combining this relation with the thermal OPE gives a crossing equation for the thermal
one-point data.  Unlike the squared OPE coefficients in a reflection-positive
identical-scalar vacuum four-point function, thermal OPE coefficients need not have a
definite sign, so standard positivity-based numerical-bootstrap methods do not apply
directly.

An early proposal to use thermal OPE consistency and periodicity to constrain CFT
data appeared in Ref.~\cite{ElShowk:2011emergent}.  The modern thermal conformal
bootstrap was developed in Ref.~\cite{Iliesiu:2018fao},
where the KMS relation was recast as a crossing equation for thermal two-point
functions.  That work also derived a thermal Lorentzian inversion formula,
building on the vacuum inversion framework \cite{CaronHuot:2017analyticity},
and developed systematic large-spin perturbation theory.  Closely related OPE-inversion methods
were applied to bosonic and fermionic vector models and their thermal gap equations in
Ref.~\cite{Petkou:2018ynm}; a quantitative application to the three-dimensional Ising
CFT followed in Ref.~\cite{Iliesiu:2018zlz}.  Subsequent developments include broken
thermal conformal Ward identities, direct KMS sum rules and Tauberian control of heavy
thermal data, numerical solutions for the critical $O(N)$ models, and extensions to
thermal line defects
\cite{Marchetto:2023xap,Marchetto:2023lsb,Barrat:2025wbi,Barrat:2024defects}.
More recent non-positivity-based approaches combine KMS, dispersion relations, and
parametrizations of the infinite OPE tail \cite{Niarchos:Deep2025}.

Thermal correlators also probe holographic dynamics, from black-hole interiors
to quasinormal spectra \cite{Grinberg:2020interior,Ceplak:2024singularity,
Dodelson:2023product,Arnaudo:2026OPEQNM}.  For two external stress tensors,
the OPE determines multi-stress data and organizes near-lightcone behavior
\cite{Karlsson:2022thermalTT,Esper:2023thermalTT}.  For external scalars,
Buri\'c, Gusev and Parnachev reconstructed double-trace contributions from
multi-stress exchange using KMS symmetry and resummation
\cite{Buric:2025kms,Buric:HolographicBootstrap2025}.  Their extension to
nonzero spatial separation determines spin-resolved thermal coefficients,
with a residual zero Matsubara mode fixed by the bulk wave equation
\cite{Buric:2026spinresolved}.

On $S^1_\beta\times S^{d-1}$ and rotating backgrounds, thermal effective
actions and spherical one-point blocks constrain high-energy spectra and
averaged OPE data \cite{Gobeil:2018thermalBlocks,Benjamin:2023thermalEFT,
Benjamin:2024kdg,Buric:2024spherical,Buric:2025HHL, Simmons-Duffin:2025qox,Mauro:2026thermalEFT}.
Fast-rotation and pp-wave limits provide complementary large-spin information
\cite{Anand:2025semiuniversality,Komargodski:2026inferno}, while rotating
partition functions can also constrain the conformal anomaly
\cite{Advant:2026Ztoa}.  Here we focus on local two-point functions on the
flat thermal cylinder.

A fundamental ingredient in our approach is the method of images, which
implements thermal periodicity by summing
Euclidean-time translates, with alternating weights for antiperiodic
correlators \cite{Fulling:1987images,Iliesiu:2018fao}.  Its modern thermal
bootstrap application combines the dispersion relation of
Ref.~\cite{Alday:2020eua} with a generalized image sum of the dispersive
contribution \cite{Barrat:2025nvu}.  Using the thermal OPE as input allows applications to interacting and
holographic CFTs \cite{Barrat:Holography2025}.  Applied block by block,
this construction gives \emph{thermal Polyakov blocks}: KMS-covariant
completions of a direct-channel thermal block.  Their generated analytic
tails and their relation to large-spin thermal data were studied in
Ref.~\cite{Barrat:2026jfg}.

In this work, we propose the \emph{thermal Polyakov bootstrap}, based on
thermal Polyakov blocks.  Each block contains its direct-channel thermal
block together with additional terms analytic at the OPE origin.  These
spurious terms admit an expansion in thermal blocks
$f_{\widehat\D,\widehat J}$ at the classical double-twist dimensions
\begin{align}
    \widehat\D=2\D_\f+2\widehat n+\widehat J,
    \qquad \widehat n\in\bb Z_{\geq0},\quad
    \widehat J\in2\bb Z_{\geq0}.
    \label{intro classical dimensions}
\end{align}
After summing over the physical spectrum, these additional spurious contributions, together with possible remainder terms, need to be canceled in order to be consistent with the physical
OPE.  This gives the central results of
this work: the bosonic and fermionic thermal Polyakov bootstrap equations
\eqref{summary bosonic Polyakov equations} and
\eqref{summary fermionic Polyakov equations}.  They express thermal
consistency as a countable linear system for the thermal OPE coefficients.
Here \emph{spurious} refers to an excess contribution to the prescribed
OPE, not to the exclusion of physical operators at double-twist dimensions.


This follows the organizing idea of the vacuum Polyakov bootstrap:
impose the crossing symmetry of the vacuum conformal four-point correlators through the building blocks and recover the
physical OPE by canceling the additional spurious terms
\cite{Polyakov:1974,Gopakumar:2016wkt,Gopakumar:2016cpb,
Gopakumar:2018polyakovMellin}.  Here KMS covariance of a thermal two-point
function plays the role of crossing symmetry of a vacuum four-point
function.  The analogy also connects our equations to analytic functionals
and dispersive CFT sum rules
\cite{Mazac:2018qmi,Carmi:2019dispersion,Caron-Huot:2020adz,
Penedones:2019nonperturbativeMellin,Trinh:2021mixedDispersive}.
Recent numerical solutions of truncated vacuum and boundary Polyakov
equations provide useful comparisons for the thermal problem
\cite{Kangshabanik:2026truncated,Kangsabanik:2026bcft}.

\subsection{A brief summary of the results}
We state the equations in the double-twist basis.  Let
$\nu=(d-2)/2$ for $d>2$, and denote the direct thermal block by
\begin{align*}
 f_{\D,J}(z,\zb)
 =(z\zb)^{\D/2-\D_\f}
 C_J^{(\nu)}\!\left(\frac{z+\zb}{2\sqrt{z\zb}}\right).
\end{align*}
The generated parts of the bosonic and fermionic Polyakov blocks have the
local expansions
\begin{align}
 P_{\D,J}(z,\zb)-f_{\D,J}(z,\zb)
 &=\sum_{\widehat n,\widehat J}
 \mathcal A_{\D,J}^{\widehat n,\widehat J}
 f_{2\D_\f+2\widehat n+\widehat J,\widehat J}(z,\zb),
 \label{summary bosonic Polyakov block}\\
 P_{\D,J}^{\rm F}(z,\zb)-f_{\D,J}(z,\zb)
 &=\sum_{\widehat n,\widehat J}
 \big(\mathcal A_{\D,J}^{\widehat n,\widehat J}\big)_{\rm F}
 f_{2\D_\f+2\widehat n+\widehat J,\widehat J}(z,\zb).
 \label{summary fermionic Polyakov block}
\end{align}
Here and below $\widehat n\geq0$ and $\widehat J=0,2,\ldots$.
The coefficients are explicit finite sums.  With
$\mathscr Z_{\rm B}(s)=\zeta(s)$,
$\mathscr Z_{\rm F}(s)=\operatorname{Li}_s(-1)$ and
$(\mathcal A)_{\rm B}=\mathcal A$, they read
\begin{align*}
\big(\mathcal A_{\D,J}^{\widehat n,\widehat J}\big)_{\rm X}
={}&\big(1+(-1)^{\widehat J}\big)
\mathscr Z_{\rm X}(\widehat J+2\widehat n+2\D_\f-\D)
\sum_{r=0}^{\widehat n}
\frac{(\widehat J+2r)\Gamma(\widehat J+r)(-\nu)_r
      (\widehat J+\nu)}
     {r!(\nu)_{\widehat J+r+1}}
\nn
&\times\sum_{k=0}^{J}
\frac{(\nu)_k(\nu)_{J-k}}{k!(J-k)!}
\binom{\frac{\D+J}{2}-\D_\f-k}
      {\widehat J+\widehat n+r}
\binom{\frac{\D-J}{2}-\D_\f+k}
      {\widehat n-r},
\qquad {\rm X}={\rm B},{\rm F}.
\end{align*}
The $(\widehat J,r)=(0,0)$ term uses
$\lim_{\widehat J\to0}\widehat J\;\Gamma(\widehat J)=1$;
the continuation prescription is specified in section~\ref{Derivation of the thermal Polyakov bootstrap equations}.
For the fermionic derivative-trace structure,
$\D_\f=\D_\ps+\tfrac12$ is the effective scalar dimension.

We regard these coefficients as an infinite matrix
$\mathcal A_\mu^\lambda$, the \emph{Polyakov matrix}.
Its columns are labeled by physical OPE operators
$\mu=(\D,J)$, and its rows by the generated classical double-twist
blocks $\lambda=(\widehat n,\widehat J)$.  Each matrix element gives
the contribution of one physical operator to one cancellation equation.

Cancellation of each generated double-twist block gives the main results, the thermal Polyakov bootstrap equations:
\begin{align}
 \boxed{\quad
 \mathcal B^{\widehat n,\widehat J}
 +\sum_{\{\D,J\}}a_{\D,J}\,
 \mathcal A_{\D,J}^{\widehat n,\widehat J}=0
 \quad}
 &\qquad\text{(bosons)},
 \label{summary bosonic Polyakov equations}\\
 \boxed{\quad
 \mathcal B_{\rm F}^{\widehat n,\widehat J}
 +\sum_{\{\D,J\}}a_{\D,J}\,
 \big(\mathcal A_{\D,J}^{\widehat n,\widehat J}\big)_{\rm F}=0
 \quad}
 &\qquad\text{(fermions)}.
 \label{summary fermionic Polyakov equations}
\end{align}
The sums run over the physical spectrum, including the unit operator with
$a_{0,0}=1$ in our normalization.  The coefficients $\mathcal B$ encode any compensating contributions from
contour terms or the interchange of infinite sums.  They vanish when the
blockwise reconstruction is complete; otherwise they must be fixed by the
reconstruction prescription and physical conditions.

Zero spatial separation offers a useful simplification.  When the
dispersive reconstruction reproduces the branch cut behavior and pole
terms prescribed by the OPE and obeys the growth conditions in
section~\ref{sec:zero-distance}, the bosonic correlator is fixed up to
one additive constant \cite{Barrat:2025nvu}.  Thus only the level-zero
zero-separation equation has a source; every positive-level equation is
source-free.  At $z=\zb=\tau$, each equation combines spins at fixed
$2\widehat n+\widehat J$, so this statement does not set the individual
spin-resolved $\mathcal B^{\widehat n,\widehat J}$ to zero.  The
antiperiodic structure has no constant ambiguity under the same assumptions.

In section \ref{sec:block-applications} we apply these equations to a set of classical thermal CFTs and find
consistent solutions.  For interacting theories, the double-twist part of
the matrix provides a simple starting point for combining known vacuum
spectral data with a large-spin tail.  Section~\ref{sec:large-spin-numerics}
explains this structure and outlines a numerical strategy for the 3D Ising
model at finite temperature, in which the large-spin equations eliminate tail coefficients before
the remaining equations constrain the low-lying thermal data.

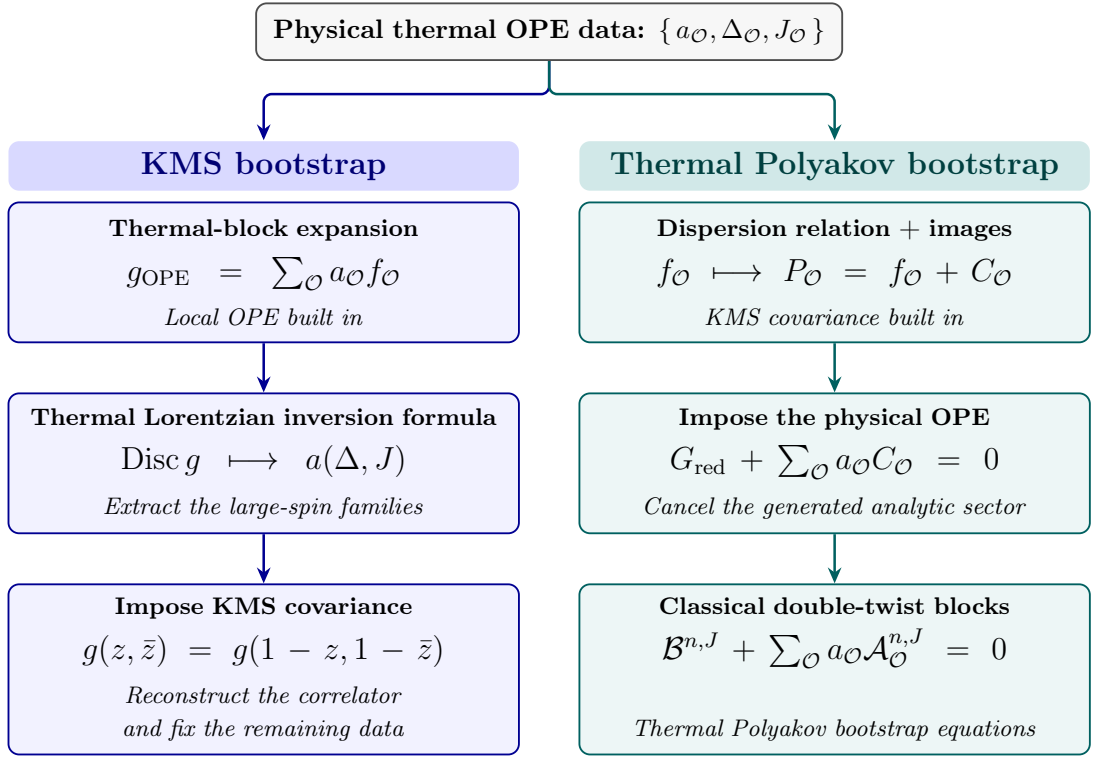
\begin{figure}[H]
    \centering
\begin{tikzpicture}[
    x=1cm,y=1cm,
    font=\fontsize{9}{11.5}\selectfont,
    box/.style={rounded corners=4pt,align=center,line width=.7pt,
                inner xsep=4.5pt,inner ysep=4pt,text width=6.40cm,
                minimum width=6.75cm,minimum height=1.85cm},
    leftbox/.style={box,draw=blue!57!black,fill=blue!5},
    rightbox/.style={box,draw=teal!75!black,fill=teal!7},
    heading/.style={box,draw=none,minimum height=.62cm,
                    font=\fontsize{12}{14}\selectfont\bfseries},
    route/.style={-{Stealth[length=2.1mm,width=1.8mm]},line width=.95pt},
    leftroute/.style={route,draw=blue!57!black},
    rightroute/.style={route,draw=teal!75!black}
]
  \node[box,draw=black!70,fill=black!3,text width=7.4cm,
        minimum width=7.75cm,minimum height=.75cm,
        font=\fontsize{10}{12}\selectfont] (data) at (0,0)
    {\textbf{Physical thermal OPE data:}
     $\{\,a_{\mathcal O}, \Delta_{\mathcal O},J_{\mathcal O}\,\}$};

  \node[heading,fill=blue!15,text=blue!45!black] (lefthead)
    at (-3.775,-1.775) {KMS bootstrap};
  \node[heading,fill=teal!18,text=teal!50!black] (righthead)
    at (3.775,-1.775) {Thermal Polyakov bootstrap};
  \draw[leftroute,rounded corners=3pt]
    (data.south) -- (0,-.825) -- (-3.775,-.825) -- (lefthead.north);
  \draw[rightroute,rounded corners=3pt]
    (data.south) -- (0,-.825) -- (3.775,-.825) -- (righthead.north);

  \node[leftbox,anchor=north] (ope) at (-3.775,-2.245)
    {\textbf{Thermal-block expansion}\\[4pt]
     {\fontsize{12}{13}\selectfont
      $g_{\mathrm{OPE}}=\sum_{\mathcal O}a_{\mathcal O}f_{\mathcal O}$}\\[4pt]
     \textit{Local OPE built in}};
  \node[rightbox,anchor=north] (images) at (3.775,-2.245)
    {\textbf{Dispersion relation + images}\\[4pt]
     {\fontsize{12}{13}\selectfont
      $f_{\mathcal O}\longmapsto P_{\mathcal O}=f_{\mathcal O}+C_{\mathcal O}$}\\[4pt]
     \textit{KMS covariance built in}};

  \node[leftbox,anchor=north] (inversion) at (-3.775,-4.770)
    {\textbf{Thermal Lorentzian inversion formula}\\[4pt]
     {\fontsize{12}{13}\selectfont
      $\mathrm{Disc}\,g\longmapsto a(\Delta,J)$}\\[4pt]
     \textit{Extract the large-spin families}};
  \node[rightbox,anchor=north] (cancel) at (3.775,-4.770)
    {\textbf{Impose the physical OPE}\\[4pt]
     {\fontsize{12}{13}\selectfont
      $G_{\mathrm{red}}+\sum_{\mathcal O}a_{\mathcal O}C_{\mathcal O}=0$}\\[4pt]
     \textit{Cancel the generated analytic sector}};
  \draw[leftroute] (ope.south) -- (inversion.north);
  \draw[rightroute] (images.south) -- (cancel.north);

  \node[leftbox,anchor=north,minimum height=2.25cm] (kms)
    at (-3.775,-7.295)
    {\textbf{Impose KMS covariance}\\[4pt]
     {\fontsize{12}{13}\selectfont
      $g(z,\bar z)=g(1-z,1-\bar z)$}\\[4pt]
     \textit{Reconstruct the correlator}\\
     \textit{and fix the remaining data}};
  \node[rightbox,anchor=north,minimum height=2.25cm] (twist)
    at (3.775,-7.295)
    {\textbf{Classical double-twist blocks}\\[4pt]
     {\fontsize{12}{13}\selectfont
      $\mathcal B^{n,J}+\sum_{\mathcal O}a_{\mathcal O}
        \mathcal A_{\mathcal O}^{n,J}=0$}\\[4pt]
     \strut\\
     \textit{Thermal Polyakov bootstrap equations}};
  \draw[leftroute] (inversion.south) -- (kms.north);
  \draw[rightroute] (cancel.south) -- (twist.north);
\end{tikzpicture}
    \caption{Two routes to thermal consistency.  The KMS bootstrap uses
    thermal inversion to reconstruct large-spin data and then imposes KMS
    crossing.  The Polyakov bootstrap builds KMS covariance into the blocks
    and imposes cancellation of the generated double-twist contributions.}
    \label{fig:summary}
\end{figure}

\section{Derivation of the thermal Polyakov bootstrap equations}
\label{Derivation of the thermal Polyakov bootstrap equations}
Our goal is to turn KMS covariance and the local OPE into explicit linear
equations for thermal CFT data at general spatial separation.  We first
construct KMS-covariant blocks using dispersion and thermal images, then
project their excess OPE contributions, i.e., the spurious terms, onto the classical double-twist
basis.  The last subsection explains how this coefficient formulation
reorganizes the usual KMS constraints. 

Set $\beta=1$ and consider the thermal two-point function
\begin{align}
    g(z,\zb)=\<\f(z,\zb)\f(0,0)\>_{S^1\times \bb R^{d-1}}
    \,,
    \label{thermal correlator definition}
\end{align}
in the coordinates
\begin{align}
    z=\t+i|\bf{x}|\,,\qquad
    \zb=\t-i|\bf{x}|
    \,.
    \label{thermal coordinate definition}
\end{align}
For $d>2$, our normalization of the direct-channel thermal OPE is
\cite{Iliesiu:2018fao}
\begin{align}
 g(z,\zb)&=\sum_{\{\D,J\}}a_{\D,J}f_{\D,J}(z,\zb),\nn
 f_{\D,J}(z,\zb)
 &= (z\zb)^{\frac{\D}{2}-\D_\f}
 C_J^{(\nu)}\!\left(\frac{z+\zb}{2\sqrt{z\zb}}\right),
 \qquad \nu=\frac{d-2}{2}.
 \label{direct thermal block normalization}
\end{align}
The identity has $f_{0,0}=(z\zb)^{-\D_\f}$ and $a_{0,0}=1$;
identical external scalars exchange only even spins.  The $d=2$ normalization
is given in section~\ref{sec:two-dimensional-examples}.
With $r=\sqrt{z\zb}$ and $\eta=(z+\zb)/(2r)$, the OPE converges for
$r<1$ \cite{Iliesiu:2018fao}.  At zero spatial separation its angular
dependence collapses:
\begin{align}
 g(\tau,\tau)
 =\sum_{\D}\left[\sum_J a_{\D,J}C_J^{(\nu)}(1)\right]
 \tau^{\D-2\D_\f}.
 \label{zero distance spin degeneracy}
\end{align}
Keeping $z$ and $\zb$ independent resolves spin.

Let $X={\rm B},{\rm F}$ label the periodic and antiperiodic scalar
structures, with $\eta_{\rm B}=1$ and $\eta_{\rm F}=-1$.  Their KMS
relations are \cite{Iliesiu:2018fao,Marchetto:2023lsb}
\begin{align}
 g_X(z+1,\zb+1)&=\eta_X\,g_X(z,\zb),\nn
 g_X(z,\zb)&=\eta_X\,g_X(1-z,1-\zb).
 \label{thermal covariance unified}
\end{align}
The second relation also uses the reflection symmetry
$g_X(-z,-\zb)=g_X(z,\zb)$ of the scalar structures considered here.
We first construct blocks satisfying these relations, then match their
sum to the physical OPE.

\subsection{Thermal dispersion relation and method of images}

The thermal dispersion relation expresses the correlator through its cut
discontinuity, with contour and subtraction terms specified below.
We apply it to a single crossed thermal block and sum its translates
around the thermal circle.  Dispersion controls the analytic continuation,
while the image sum implements KMS covariance.

\begin{figure}[H]
    \centering
    \begin{tikzpicture}[
    x=2.20cm,
    y=2.20cm,
    font=\small,
    axis/.style={draw=black!60,line width=0.55pt,-{Stealth[length=2mm]}},
    euclidean/.style={draw=blue!68!black,line width=1.25pt},
    cut/.style={draw=red!72!black,line width=1.15pt,
      decorate,decoration={zigzag,segment length=4.2pt,amplitude=1.8pt}},
    branch/.style={draw=red!72!black,fill=orange!75,line width=0.55pt}
]
  \def\rin{0.55}
  \def\rout{1.82}

  \draw[axis] (-2.55,0)--(2.62,0)
    node[above left=1pt] {$\operatorname{Re}w$};
  \draw[axis] (0,-1.42)--(0,1.48)
    node[below right=1pt] {$\operatorname{Im}w$};

  \draw[euclidean] (0,0) circle (1);
  \node[align=center,text=blue!68!black] at (1.18,1.08)
    {Euclidean locus\\[-1pt]$|w|=1$};

  \draw[cut] (-2.53,0)--(-\rout,0);
  \draw[cut] (-\rin,0)--(-0.055,0);
  \draw[cut] (0.055,0)--(\rin,0);
  \draw[cut] (\rout,0)--(2.53,0);

  \foreach \x in {-\rout,-\rin,0,\rin,\rout}
    \draw[branch] (\x,0) circle (0.045);

  \node[below=5pt] at (-\rout,0) {$-r^{-1}$};
  \node[below=5pt] at (-\rin,0) {$-r$};
  \node[below=5pt] at (0,0) {$0$};
  \node[below=5pt] at (\rin,0) {$r$};
  \node[below=5pt] at (\rout,0) {$r^{-1}$};

\end{tikzpicture}
    \caption{Analytic structure of
    $\mathfrak g_r(w)=g(rw,r/w)$ at fixed $0<r<1$.  Euclidean kinematics
    lie on $|w|=1$, while the red zigzag lines denote the cuts
    $[-r,0)\cup(0,r]$ and
    $(-\infty,-r^{-1}]\cup[r^{-1},\infty)$.}
    \label{fig:contour4xneq0}
\end{figure}
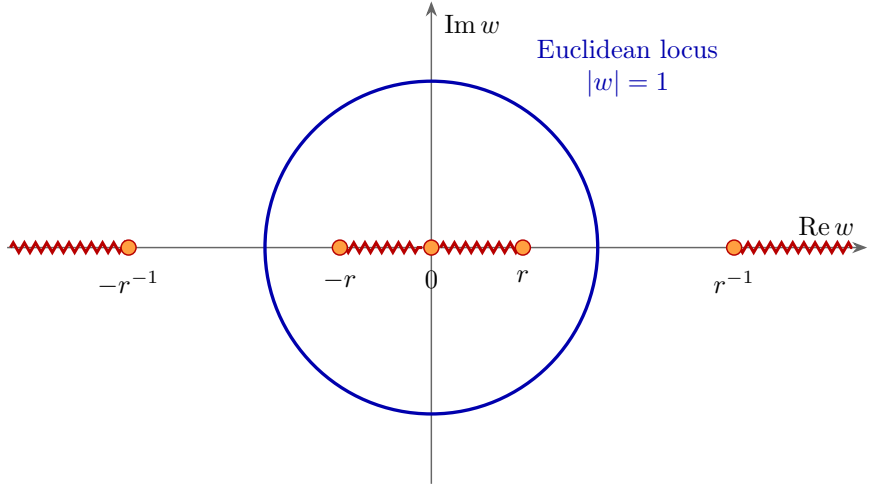

For the thermal dispersion relation \cite{Alday:2020eua}, fix $0<r<1$
on the Euclidean sheet and set $\mathfrak g_r(w)=g(rw,r/w)$.
The OPE gives analyticity in $r<|w|<r^{-1}$
\cite{Iliesiu:2018fao}, with Euclidean locus $|w|=1$.
Exchange and reflection symmetry imply
$\mathfrak g_r(w)=\mathfrak g_r(1/w)=\mathfrak g_r(-w)$.
Figure~\ref{fig:contour4xneq0} shows the analytic structure of
$\mathfrak g_r(w)$ in the complex $w$-plane.  As in vacuum dispersion
relations \cite{Carmi:2019dispersion,Caron-Huot:2020adz}, deforming the
Cauchy contour expresses the thermal correlator as integrals along the two
sides of these cuts, together with integrals around branch points or poles
and any surviving contribution from the contour at infinity.

We start from the contour integral identity
\begin{align*}
 g(z,\zb)&=\oint_{\mathcal C_0}\frac{{\rm d}w}{2\pi i}
 K_0(z,\zb;w)\,\mathfrak g_r(w),&
 K_0(z,\zb;w)&=\frac{1}{w(1-w^2z/\zb)}.
\end{align*}
Here $\mathcal C_0$ consists of clockwise contours around
$w=\pm\sqrt{\zb/z}$, enclosing no cut or origin; each kernel residue is
$-1/2$.  Deforming and folding with the two symmetries gives
\begin{align}
g_{\rm{disc}}(z,\zb)
&=\int_0^r\frac{{\rm d}w}{\pi i}\,
\mathcal K(z,\zb;w)\,\Disc_z[g(r/w,rw)],\nn
\mathcal K(z,\zb;w)&=\frac1w\left(\frac1{1-w^2z/\zb}
+\frac1{1-w^2\zb/z}-1\right)
\,,\label{dispersion}
\end{align}
where we use the unnormalized discontinuity in the first coordinate,
\begin{align}
\Disc[F(z)]=F(z+i0)-F(z-i0)
\,.
\label{thermal discontinuity convention}
\end{align}
In the folded integral we use the equivalent parametrization
$\mathfrak g_r(w)=g(r/w,rw)$.  Boundary values are taken on the same
Euclidean sheet; $z=r/w$ reverses the orientation, so $\Disc_z$ and
$\Disc_w$ have opposite signs on this segment.  The full relation is
\begin{align}
 g(z,\zb)=g_{\rm{disc}}(z,\zb)+g_{\rm{hom}}(z,\zb).
 \label{dispersion plus arc}
\end{align}
The cut integral is understood by analytic continuation, with local
subtractions near $w=0$ and $w=r$ when necessary.  The term $g_{\rm{hom}}$
contains the contour contributions and subtraction terms not included in
$g_{\rm{disc}}$.
Appendix~\ref{sec:endpoint-prescription} specifies this prescription and
the growth conditions at infinity.

Bosonic KMS crossing gives
\begin{align}
g(z,\zb)=g(1-z,1-\zb)
=\sum_{\{\D,J\}}a_{\D,J}f_{\D,J}(1-z,1-\zb)
\,.
\label{crossed thermal OPE}
\end{align}
For a single crossed block, define the continued cut contribution
\begin{align}
\cal P_{\D,J}(z,\zb)=\int_0^r
\frac{{\rm d}w}{\pi i}\,\mathcal K(z,\zb;w)
\Disc_z[f_{\D,J}(1-r/w,1-rw)]
\, .\label{continued crossed block}
\end{align}
The monomial integrals are analytically continued from a domain where
they converge at both $w=0$ and $w=r$, with the small-contour terms
evaluated in the same prescription
(appendix~\ref{sec:endpoint-prescription}).  This prescription yields
\begin{align}
 \cal P_{\D,J}(z,\zb)
 =f_{\D,J}(1-z,1-\zb)+f_{\D,J}(1+z,1+\zb).
 \label{continued two image identity}
\end{align}

Summing translates restores KMS covariance.  Performing the OPE sum
before the image sum gives the dispersive representation of
Refs.~\cite{Barrat:2025nvu,Barrat:2026jfg}:
\begin{align}
g(z,\zb)=\frac 1 2\sum_{m=-\infty}^{\infty}
\left[\sum_{\{\D,J\}}a_{\D,J}\cal P_{\D,J}(z+m,\zb+m)\right]_{\rm{reg}}
+G_1(z,\zb)
\,.\label{GMI}
\end{align}
The KMS-invariant remainder $G_1$ contains the contour and subtraction
terms from \eqref{dispersion plus arc} not already included in the
block integrals.  It also accounts for any difference between integrating
the full correlator and integrating its OPE term by term, in the same
regularization prescription.  The full correlator must have the prescribed OPE,
singularities and clustering limit
$g\to\<\f\>^2_{S^1\times\bb R^{d-1}}$ as $|\mathbf x|\to\infty$
\cite{Barrat:2025nvu,Barrat:Holography2025};
section~\ref{sec:ON-epsilon-block-example} gives an example.
Define the thermal Polyakov block by
\begin{align}
P_{\D,J}(z,\zb)=\frac 1 2\sum_{m=-\infty}^{\infty}
\cal P_{\D,J}(z+m,\zb+m)
\,,\label{Polyakov block}
\end{align}
The image sum is first defined in an absolutely convergent domain and
then continued meromorphically.  Integer shifts of the summation index
preserve this prescription.  Reindexing \eqref{continued two image identity}
then gives \cite{Barrat:2026jfg}
\begin{align}
P_{\D,J}(z,\zb)
=\left[\sum_{m=-\infty}^{\infty}
f_{\D,J}(z+m,\zb+m)\right]_{\rm{reg}},
\label{ordinary image representation}
\end{align}
where the factor $1/2$ in \eqref{Polyakov block} removes double counting.
Euclidean branches have
$R_m=\sqrt{(\tau+m)^2+|\mathbf x|^2}>0$ and, for even $J$ near the OPE point,
 $f(z-m,\zb-m)=f(m-z,m-\zb)$.
The periodicity relation $P_{\D,J}(z+1,\zb+1)=P_{\D,J}(z,\zb)$
follows by shifting the summation index.
Interchanging the OPE and image sums in \eqref{GMI} gives
\begin{align}
g(z,\zb)=\sum_{\{\D,J\}}a_{\D,J}P_{\D,J}(z,\zb)+G_{\rm{red}}(z,\zb)
\,,\label{interchange sums}
\end{align}
Here $G_{\rm{red}}=g_{\rm{arcs}}+G_{\rm{reorder}}$ denotes the full
reconstruction remainder.
The term $g_{\rm{arcs}}$ collects contour and subtraction terms, while
$G_{\rm{reorder}}$ accounts for differences caused by moving the OPE sum
through the contour integral and the image sum.  Its Taylor coefficients
vanish under the sufficient conditions in
appendix~\ref{sec:summation-prescription}.  Contour and subtraction terms
must still be evaluated in the chosen prescription.  For example, the scalar reconstruction of
Ref.~\cite{Buric:2026spinresolved} fixes its residual zero Matsubara mode
through the bulk wave equation.

For antiperiodic scalar structures extracted from fermion correlators, the
crossing sign in \eqref{thermal covariance unified} and alternating images give
\begin{align}
g(z,\zb)=-\frac 1 2\sum_{m=-\infty}^{\infty}(-1)^m
\left[\sum_{\{\D,J\}}a_{\D,J}\cal P_{\D,J}(z+m,\zb+m)\right]_{\rm{reg}}
+G_{1,\rm F}(z,\zb)
\,.\label{GMI fermion}
\end{align}
The remainder $G_{1,\rm F}$ is defined by the same contour and summation
prescription.  Define
\begin{align}
P_{\D,J}^{\rm F}(z,\zb)
&=-\frac 1 2\sum_{m=-\infty}^{\infty}(-1)^m
\cal P_{\D,J}(z+m,\zb+m)
\nn
&=\left[\sum_{m=-\infty}^{\infty}(-1)^m
f_{\D,J}(z+m,\zb+m)\right]_{\rm{reg}}
\,.\label{fermion Polyakov block}
\end{align}
Shifting the summation index gives $P_{\D,J}^{\rm F}(z+1,\zb+1)=-P_{\D,J}^{\rm F}(z,\zb)$;
the minus sign from either translate in \eqref{continued two image identity}
explains the prefactor $-1/2$.  Together with
$f_{\D,J}(-z,-\zb)=f_{\D,J}(z,\zb)$ for even $J$, this proves
\begin{align}
P_{\D,J}(1-z,1-\zb)&=P_{\D,J}(z,\zb),&
P_{\D,J}^{\rm F}(1-z,1-\zb)&=-P_{\D,J}^{\rm F}(z,\zb).
\label{Polyakov block KMS relations}
\end{align}
Alternating weights do not by themselves justify exchanging infinite sums.
Denoting the full antiperiodic reconstruction remainder by
$G_{\mathrm{red},\mathrm F}$ gives
\begin{align}
g(z,\zb)=\sum_{\{\D,J\}}a_{\D,J}P_{\D,J}^{\rm F}(z,\zb)+G_{\mathrm{red},\mathrm F}(z,\zb)
\,.\label{fermion Polyakov block expansion}
\end{align}
For either choice of statistics, each completed block contains its designated physical thermal
block and an analytic image tail.  Matching the prescribed OPE requires the
net excess from these tails and the compensating term to vanish, including
when a generated block coincides with a physical operator.

\subsection{Thermal Polyakov bootstrap equations in the double-twist basis}
\label{sec:standard-thermal-polyakov}
\label{Polyakov-type bootstrap equations}

We now extract the bootstrap equations from the KMS-covariant
representation. The spurious terms in the KMS-covariant expansion can be expressed in
different bases.  Here we adopt the double-twist basis, which resolves the generated analytic
terms by radial level and spin, giving one equation for each basis
element.  It also makes the classical double-twist submatrix diagonal,
which will be useful for solving the equations.  We give the closed
kernels for both statistics and then state the equivalent monomial form.

The nonzero images generate terms analytic at the OPE origin.  At even
degree $L=2\widehat n+\widehat J$, the $L/2+1$ polynomials
\begin{align}
 f_{2\D_\f+2\widehat n+\widehat J,\widehat J}(z,\zb)
 =\sum_{p=0}^{\widehat J}
 \frac{(\nu)_p(\nu)_{\widehat J-p}}{p!(\widehat J-p)!}
 z^{\widehat n+p}\zb^{\widehat n+\widehat J-p},
 \qquad \widehat J=0,2,\ldots,L,
 \label{double twist homogeneous polynomial}
\end{align}
form a basis of the symmetric homogeneous polynomials.  Following the block
decomposition of Ref.~\cite{Barrat:2026jfg}, we expand the image tails and
the same remainder $G_{\rm{red}}$ defined in \eqref{interchange sums}.
For the latter expansion, we require analyticity at the OPE origin and a
summation prescription that permits Taylor coefficient extraction:

\begin{align}
 P_{\D,J}-f_{\D,J}
 &=\sum_{\widehat n,\widehat J}
 \mathcal A_{\D,J}^{\widehat n,\widehat J}
 f_{2\D_\f+2\widehat n+\widehat J,\widehat J},
 \label{double twist Polyakov decomposition}\\
 G_{\rm{red}}
 &=\sum_{\widehat n,\widehat J}
 \mathcal B^{\widehat n,\widehat J}
 f_{2\D_\f+2\widehat n+\widehat J,\widehat J}.
 \label{block source expansion}
\end{align}
Here $\widehat n\geq0$ and $\widehat J$ is even.  The hatted labels specify
the generated basis, not an assumed physical spectrum.
Analyticity and the basis property above make the expansion of
$G_{\rm{red}}$ unique.  A finite limit at the origin alone would not
suffice for this power-series expansion.

Matching the physical OPE gives
\begin{align}
 G_{\rm{red}}+\sum_{\{\D,J\}}a_{\D,J}(P_{\D,J}-f_{\D,J})=0,
 \label{analytic cancellation identity}
\end{align}
hence the bosonic thermal Polyakov bootstrap equations
\begin{align}
 \boxed{\quad
 \mathcal B^{\widehat n,\widehat J}
 +\sum_{\{\D,J\}}a_{\D,J}
 \mathcal A_{\D,J}^{\widehat n,\widehat J}=0
 \quad}.
 \label{new Polyakov equation}
\end{align}
The sum includes $a_{0,0}=1$, and
$\mathcal A_{\mathds1}\equiv\mathcal A_{0,0}$.  Note that a generated block may
coincide with a physical operator; the equation cancels its excess
contribution, not its physical OPE coefficient.

For $d>2$, $\nu=(d-2)/2$, the explicit kernel is 
\begin{align}
 \mathcal A_{\D,J}^{\widehat n,\widehat J}
 &=\big(1+(-1)^{\widehat J}\big)
 \zeta\!\left(s_{\widehat n\widehat J}(\D)\right)
 \mathcal Q_{\D,J}^{\widehat n,\widehat J},
 \label{unphysical coefficient new}\\
 s_{\widehat n\widehat J}(\D)
 &=\widehat J+2\widehat n+2\D_\f-\D,
 \label{block image exponent}\\
 \mathcal Q_{\D,J}^{\widehat n,\widehat J}
 &=\sum_{r=0}^{\widehat n}
 \frac{(\widehat J+2r)\Gamma(\widehat J+r)(-\nu)_r
       (\widehat J+\nu)}
      {r!(\nu)_{\widehat J+r+1}}
 \sum_{k=0}^{J}\frac{(\nu)_k(\nu)_{J-k}}{k!(J-k)!}
 \nn
 &\hspace{1cm}\times
 \binom{\frac{\D+J}{2}-\D_\f-k}{\widehat J+\widehat n+r}
 \binom{\frac{\D-J}{2}-\D_\f+k}{\widehat n-r}.
 \label{double twist kinematic kernel}
\end{align}
The $k$-sum expands the input Gegenbauer polynomial; the $r$-sum projects
onto the output spin.  Their derivation is given in
appendix~\ref{app:kernel-projections}.

For the antiperiodic scalar structure, replace the image factor by
$\operatorname{Li}_s(-1)$:
\begin{gather}
 \big(\mathcal A_{\D,J}^{\widehat n,\widehat J}\big)_{\rm F}
 =\big(1+(-1)^{\widehat J}\big)
 \operatorname{Li}_{s_{\widehat n\widehat J}(\D)}(-1)
 \mathcal Q_{\D,J}^{\widehat n,\widehat J},
 \label{unphysical coefficient fermion new}\\
 \boxed{\quad
 \mathcal B_{\rm F}^{\widehat n,\widehat J}
 +\sum_{\{\D,J\}}a_{\D,J}
 \big(\mathcal A_{\D,J}^{\widehat n,\widehat J}\big)_{\rm F}=0
 \quad}.
 \label{new Polyakov equation fermion}
\end{gather}
For the derivative-trace construction in section~\ref{sec:fermionic-generalized-free},
$\D_\f=\D_\ps+\tfrac12$.  Both systems are affine after identity normalization;
they are source-free only when the corresponding $\mathcal B$ vanishes.

\paragraph{Classical double-twist columns.}
Set $\D=2\D_\f+2m+J$, $L_0=2m+J$ and $L=2\widehat n+\widehat J$.
The two upper binomial entries in \eqref{double twist kinematic kernel}
are nonnegative integers with sum $L_0$, whereas the lower entries sum
to $L$.  Thus
\begin{align}
 L&>L_0:\quad \mathcal Q=0,\nn
 L&<L_0:\quad \zeta(L-L_0)=\operatorname{Li}_{L-L_0}(-1)=0,
 \nn
 L&=L_0:\quad
 \mathcal Q_{2\D_\f+2m+J,J}^{\widehat n,\widehat J}
 =\delta_{m,\widehat n}\delta_{J,\widehat J}.
 \label{double twist self kernel}
\end{align}
The last equality follows from uniqueness of the basis expansion.  Since
$2\zeta(0)=2\operatorname{Li}_0(-1)=-1$,
\begin{align}
 \boxed{\quad
 \mathcal A_{2\D_\f+2m+J,J}^{\widehat n,\widehat J}
 =\left(\mathcal A_{2\D_\f+2m+J,J}^{\widehat n,\widehat J}\right)_{\rm F}
 =-\delta_{m,\widehat n}\delta_{J,\widehat J}
 \quad}.
 \label{double twist diagonal kernel}
\end{align}
This full submatrix is $-I$ in the continued image prescription.  It
provides the reference matrix for the numerical construction in
section~\ref{sec:ising-numerics}. 

A literal image sum of a polynomial does not converge, so the above
formulas should be interpreted as analytic continuations.

\paragraph{Equivalent monomial equations.}
For coefficient extraction or zero-separation projections, one may instead use
\begin{align}
 C_{\D,J}\equiv P_{\D,J}-f_{\D,J}
 &=\sum_{\a,\b\geq0}A_{\D,J}^{\a,\b}z^\a\zb^\b,
 &G_{\rm{red}}&=\sum_{\a,\b\geq0}B^{\a,\b}z^\a\zb^\b,
 \label{monomial Polyakov decomposition}\\
 s_{\a\b}(\D)&=\a+\b+2\D_\f-\D.
 \label{monomial image exponent}
\end{align}
With $H_{\D,J}^{\a,\b}$ given explicitly in
appendix~\ref{app:monomial-kernel},
\begin{align}
 A_{\D,J}^{\a,\b}
 &=\big(1+(-1)^{\a+\b}\big)\zeta(s_{\a\b})H_{\D,J}^{\a,\b},
 \label{unphysical coefficient}\\
 \big(A_{\D,J}^{\a,\b}\big)_{\rm F}
 &=\big(1+(-1)^{\a+\b}\big)\operatorname{Li}_{s_{\a\b}}(-1)H_{\D,J}^{\a,\b}.
 \label{unphysical coefficient fermion}
\end{align}
The corresponding equations are
\begin{align}
 B^{\a,\b}+\sum_{\{\D,J\}}a_{\D,J}A_{\D,J}^{\a,\b}&=0,
 \label{Polyakov bootstrap}\\
 B_{\rm F}^{\a,\b}+\sum_{\{\D,J\}}a_{\D,J}
 \big(A_{\D,J}^{\a,\b}\big)_{\rm F}&=0.
 \label{Polyakov bootstrap fermion}
\end{align}
For even input spin,
\begin{align}
 A_{\D,J}^{\a,\b}=A_{\D,J}^{\b,\a},\qquad
 A_{\D,J}^{\a,\b}=0\quad(\a+\b\ {\rm odd}),
 \label{monomial kernel symmetries}
\end{align}
with the same symmetries for the fermionic coefficients and sources.
At fixed even $L$, the $L/2+1$ rows with $\a\leq\b$, $\a+\b=L$ are related
to the double-twist rows by
\begin{align}
 A_{\D,J}^{\a,\b}
 &=\sum_{\widehat n,\widehat J}
 M_{\widehat n,\widehat J}^{\a,\b}\mathcal A_{\D,J}^{\widehat n,\widehat J},
 \nn
 M_{\widehat n,\widehat J}^{\a,\b}
 &=\delta_{\a+\b,\,2\widehat n+\widehat J}
 \frac{(\nu)_{\a-\widehat n}(\nu)_{\b-\widehat n}}
 {(\a-\widehat n)!(\b-\widehat n)!}.
 \label{basis change matrix}
\end{align}
Entries with a negative factorial argument vanish.  The sources transform
by the same matrix, which is triangular and invertible on this symmetric
subspace.  Thus the two forms impose identical constraints.
The $d=2$ normalization is given in section~\ref{sec:two-dimensional-examples}.

\subsection{Relation to the KMS crossing bootstrap}
\label{sec:kms-comparison}
\label{sec:vacuum-comparison}

The vacuum comparison provides an instructive benchmark.  Caron-Huot,
Maz\'a\v{c}, Rastelli and Simmons-Duffin \cite{Caron-Huot:2020adz}
related position-space dispersion, double-twist functionals and the Mellin-space
Polyakov bootstrap through explicit contour operators.  Their
Polyakov--Regge expansion and functional relations hold on specified
analytic function spaces, with Regge bounds and, when necessary,
subtractions.  In particular, Ref.~\cite{Caron-Huot:2020adz} showed that the
Mellin Polyakov conditions are identical to the actions of functionals on
the ordinary crossing equations.
For the thermal problem, we can exhibit an explicit forward relation
between the Polyakov rows and KMS crossing; the converse also depends on
the reconstruction ambiguity.

\paragraph{From Polyakov coefficients to KMS crossing.}
Using the statistics sign $\eta_X$ defined in
\eqref{thermal covariance unified}, define
\begin{equation}
    \begin{split}
        (\mathsf K_{\rm X}h)(z,\zb)
 &=\eta_Xh(1-z,1-\zb), \\
 F_{\rm X}[h]&=(1-\mathsf K_{\rm X})h.
 \label{thermal crossing operator}
    \end{split}
\end{equation}
The usual thermal bootstrap imposes
$F_{\rm X}[g_{\rm{OPE}}]=\sum_{\mathcal O}a_{\mathcal O}
F_{\rm X}[f_{\mathcal O}]=0$ \cite{Iliesiu:2018fao}.
Let $\lambda=(\widehat n,\widehat J)$ and
$H_\lambda=f_{2\D_\f+2\widehat n+\widehat J,\widehat J}$.
Each Polyakov block obeys KMS covariance,
\begin{equation}
F_X[P_{\mathcal O}^{\rm X}]=0,    
\end{equation}
and its expansion gives
\begin{align}
 P_{\mathcal O}^{\rm X}
 &=f_{\mathcal O}+\sum_\lambda
 (\mathcal A_{\mathcal O}^{\lambda})_{\rm X}H_\lambda,
 \label{redundant reconstruction map}\\
 F_{\rm X}[f_{\mathcal O}]
 &=-\sum_\lambda(\mathcal A_{\mathcal O}^{\lambda})_{\rm X}
 F_{\rm X}[H_\lambda].
 \label{thermal crossing vector expansion}
\end{align}
This is a change of representation of the crossing vectors, not an
assumption about the physical spectrum.  Termwise reflection requires
convergence at both reflected points, or a common continuation that
justifies this operation.

Write $G_{\mathrm{red},\mathrm B}=G_{\rm{red}}$ and include the full
KMS-covariant remainder $G_{\mathrm{red},X}$ in
$g_{\rm{rec}}^{\rm X}=G_{\mathrm{red},X}+\sum_{\mathcal O}a_{\mathcal O}
P_{\mathcal O}^{\rm X}$.  If the full difference is analytic at the OPE
origin, its convergent expansion is
\begin{align}
 R_{\rm X}&\equiv g_{\rm{rec}}^{\rm X}-g_{\rm{OPE}}
 =\sum_\lambda e_\lambda^{\rm X}H_\lambda,\nn
 e_\lambda^{\rm X}
 &=\mathcal B_{\rm X}^\lambda+
 \sum_{\mathcal O}a_{\mathcal O}
 (\mathcal A_{\mathcal O}^{\lambda})_{\rm X}.
 \label{KMS Polyakov comparison}
\end{align}
KMS covariance of $g_{\rm{rec}}^{\rm X}$ then implies the function identity
\begin{align}
 F_{\rm X}[g_{\rm{OPE}}]
 &=-F_{\rm X}[R_{\rm X}]
 =-\sum_\lambda e_\lambda^{\rm X}F_{\rm X}[H_\lambda].
 \label{thermal KMS residual map}
\end{align}
The last equality additionally requires justified summation and
continuation.  Thus vanishing of every Polyakov row sets $R_{\rm X}=0$
locally, and hence enforces both OPE matching and KMS throughout the
connected common analytic domain.

\paragraph{The converse and the subtraction sector.}
KMS alone implies $F_{\rm X}[R_{\rm X}]=0$, not $R_{\rm X}=0$:
the map in \eqref{thermal KMS residual map} has an invariant kernel.
The lowest bosonic double-twist block illustrates this directly:
\begin{align}
 H_{0,0}=1,\qquad F_{\rm B}[1]=0,\qquad
 P_{2\D_\f,0}=1+2\zeta(0)=0,\qquad
 \mathcal A_{2\D_\f,0}^{0,0}=-1.
 \label{thermal constant crossing kernel}
\end{align}
Crossing cannot fix this constant.  The level-zero Polyakov
equation includes it through $\mathcal B^{0,0}$.  Consequently, the
coefficient action
$\widehat\pi_\lambda[P_{\mathcal O}^{\rm X}-f_{\mathcal O}]
=(\mathcal A_{\mathcal O}^{\lambda})_{\rm X}$ is not, by itself, the
action of a functional on the unrestricted KMS crossing vector.

To obtain the converse, the dispersion prescription must reproduce the
same branch cut discontinuities and pole terms, satisfy the required growth
bounds, and fix the remaining homogeneous terms independently.  When
these data give a unique reconstruction and the spectral interchanges
are valid, a KMS solution also satisfies the Polyakov hierarchy.
At zero separation, section~\ref{sec:zero-distance} reduces the bosonic
ambiguity to a constant.  A general inverse of
\eqref{thermal KMS residual map}, analogous to the vacuum functional
construction, is not established here.

In both the thermal Polyakov bootstrap and the KMS crossing bootstrap, the practical reorganization of the thermal data is nevertheless explicit.  Thermal inversion
reconstructs families before the remaining data are constrained by KMS,
implemented at sampled points in Ref.~\cite{Iliesiu:2018zlz}.  The Polyakov
formulation places family reconstruction and the remaining OPE constraints
in one coefficient matrix, the Polyakov matrix.
In section~\ref{sec:ising-numerics}, we will use this structure without requiring
positivity or assuming that a finite truncation establishes full crossing.

\section{Thermal Polyakov bootstrap equations with zero spatial separation}
\label{The Polyakov-type thermal bootstrap at zero-spatial distance}
\label{sec:zero-distance}
At zero spatial separation, dispersion has a specific advantage:
once the reconstruction reproduces the branch cut behavior and pole
terms prescribed by the OPE and satisfies the growth bound at imaginary
infinity, the bosonic two-point function is determined up to one additive
constant.  The zero-separation Polyakov equations therefore have no
undetermined source at positive level, unlike the general spin-resolved
system.  We make the assumptions and this simplification explicit below.

The zero-separation KMS bootstrap was developed in
Ref.~\cite{Marchetto:2023lsb}, where the contribution of the higher part
of the spectrum to the thermal correlator is approximated using
generalized free field theory.  Barrat et al.\ derived its dispersive
image blocks and constant reconstruction ambiguity \cite{Barrat:2025nvu}.
The restriction $z=\zb=\tau$, $0<\tau<1$, gives spin-weighted sums of the
equations in section~2 and explicit one-variable kernels.
Appendix~\ref{app:zero-distance} gives detailed benchmarks.

\subsection{Zero-separation kernels and loss of spin resolution}

We use the statistics sign $\eta_X$ defined in
\eqref{thermal covariance unified}, and set $s_\D=2\D_\f-\D$.
The thermal OPE at zero spatial separation becomes
\begin{align}
 g_X(\tau)&\equiv g_X(\tau,\tau)
 =\sum_\D a_\D^X\,\tau^{-s_\D},\label{thermal block expansion x=0}\\
 a_\D^X&=\sum_J a_{\D,J}^X C_J^{(\nu)}(1),\qquad
 C_J^{(\nu)}(1)=\frac{(2\nu)_J}{J!}.
 \nonumber
\end{align}
Here $\nu=(d-2)/2$ and $d>2$; the two-dimensional blocks of
section~\ref{sec:two-dimensional-examples} have unit weight at zero spatial separation.
This restriction of the $d$-dimensional OPE no longer resolves spin.

On the interval $0<\tau<1$, the image completion can be summed explicitly:
\begin{align}
 P_\D^{\rm B}(\tau)
 &=\zeta_H(s_\D,\tau)+\zeta_H(s_\D,1-\tau),\nn
 P_\D^{\rm F}(\tau)
 &=\Phi(-1,s_\D,\tau)-\Phi(-1,s_\D,1-\tau).
 \label{zero distance exact blocks}
\end{align}
The bosonic block is the one-variable image block of
Ref.~\cite{Barrat:2025nvu}.  The alternating sum uses the Lerch transcendent,
\begin{align*}
 \Phi(w,s,a)&=\sum_{m\geq0}\frac{w^m}{(m+a)^s},\\
 \Phi(-1,s,a)&=2^{-s}
 \left[\zeta_H(s,a/2)-\zeta_H(s,(a+1)/2)\right].
\end{align*}
After continuation from a convergent domain in $s$, these blocks obey
$P_\D^X(1-\tau)=\eta_X P_\D^X(\tau)$.

Separating the direct block exposes an even analytic tail,
\begin{align}
 C_\D^X(\tau)&\equiv P_\D^X(\tau)-\tau^{-s_\D}
 =\sum_{m=1}^\infty\eta_X^m
 \left[(m+\tau)^{-s_\D}+(m-\tau)^{-s_\D}\right]
 =\sum_{L=0}^\infty (A_\D^L)_X\,\tau^L,\nn
 (A_\D^L)_X&=
 \frac{1+(-1)^L}{L!}(s_\D)_L\,\mathscr Z_X(s_\D+L),  \label{unphysical coefficient x=0}\\
 \mathscr Z_{\rm B}(u)&=\zeta(u),\quad
 \mathscr Z_{\rm F}(u)=\operatorname{Li}_u(-1). \nonumber
\end{align}
The zeta poles require
a subtraction prescription (appendix~\ref{app:ON-zero-distance}).
At $s_\D=-2n$, both image blocks vanish and $(A_\D^L)_X=-\delta_{L,2n}$:
the equations reconstruct, rather than exclude, physical analytic OPE terms.

The relation to the two-variable kernels is exact in a common prescription:
\begin{align}
 P_{\D,J}^X(\tau,\tau)&=C_J^{(\nu)}(1)P_\D^X(\tau),\nn
 \sum_{\alpha+\beta=L}(A_{\D,J}^{\alpha,\beta})_X
 &=C_J^{(\nu)}(1)(A_\D^L)_X
 =\sum_{\substack{\widehat J=0,2,\ldots,L}}
 C_{\widehat J}^{(\nu)}(1)
 (\mathcal A_{\D,J}^{(L-\widehat J)/2,\widehat J})_X,
 \qquad L\ {\rm{even}}.
 \label{zero distance kernel projection}
\end{align}
At level $L=2q$, this restriction gives one linear combination of the
$q+1$ spin-resolved equations.  These zero-separation equations are
necessary but do not determine the full spin-resolved system.

\subsection{Constant ambiguity and reduced bootstrap equations}
\label{sec:zero-distance-source}
We present the thermal Polyakov bootstrap equations at zero spatial
separation.  Their simplification in this limit follows from a
one-variable reconstruction theorem
\cite{Barrat:2025nvu}.  

Let the correlation function $g_X$ be holomorphic in
$0<\operatorname{Re}u<1$, set $\tau_*=1/2$ and let
$\Gamma_{\varepsilon,R}$ be the counterclockwise boundary of
$\varepsilon<\operatorname{Re}u<1-\varepsilon$, $|\operatorname{Im}u|<R$,
containing $\tau$ and $\tau_*$.  The once-subtracted Cauchy identity is
\begin{align}
 g_X(\tau)=g_X(\tau_*)+
 \frac{\tau-\tau_*}{2\pi i}
 \oint_{\Gamma_{\varepsilon,R}}
 \frac{g_X(u)\,\mathrm du}{(u-\tau_*)(u-\tau)}.
 \label{dispersion x=0}
\end{align}
Deform the contour to both sides of the cuts, including the integrals
around branch points or poles and along the horizontal segments before
taking the limits $\varepsilon\to0$ and $R\to\infty$.  The zero-separation limit inherits cuts in both
imaginary directions.

Let $D_X$ reproduce the cut discontinuities of $g_X$ and its local
branch-point behavior, together with the principal parts at any poles
(the negative-power terms in their Laurent expansions).  The difference
$H_X=g_X-D_X$ then extends to an entire function.  If $H_X$ has the required thermal (anti)periodicity and a
uniform polynomial bound at imaginary infinity in a fundamental strip,
translations extend the bound globally.  Cauchy's estimates make $H_X$
a polynomial; periodicity forces it to be constant, while
antiperiodicity forces it to vanish.  Pole terms must be included:
ordinary cut discontinuities alone do not detect isolated poles.
Without the growth bound, entire functions such as $\cos(2\pi\tau)$
would remain undetermined.

In a common spectral and image-summation prescription that preserves
this analytic class, write
\begin{align}
 g_X(\tau)=\sum_\D a_\D^X P_\D^X(\tau)+K_X(\tau),
 \label{Polyakov block expansion x=0}
\end{align}
Here $K_X(\tau)=G_{\mathrm{red},X}(\tau,\tau)$ is the restriction of the
full reconstruction remainder.  Under the assumptions above,
\begin{align}
 K_{\rm B}(\tau)=\kappa,\qquad K_{\rm F}(\tau)=0.
 \label{zero distance bounded remainder}
\end{align}
The constant $\kappa$ depends on the subtraction prescription; it is
neither $g_X(\tau_*)$ nor, in general, the physical OPE constant.
The conclusion requires the full block sum to preserve the reconstruction
assumptions: termwise continuation does not by itself justify exchanging
infinite OPE, contour and image sums \cite{Barrat:2026jfg}.

Denote the zero-separation source by
\begin{align}
 b_X^L&=\sum_{\alpha+\beta=L}B_X^{\alpha,\beta},&
 b_X^L+\sum_\D a_\D^X(A_\D^L)_X&=0,
 \qquad L=0,2,4,\ldots .
 \label{zero distance sourced equations}
\end{align}
Equivalently, $b_X^L$ is the
$C_{\widehat J}^{(\nu)}(1)$-weighted sum of double-twist sources at
$2\widehat n+\widehat J=L$.  Equation~\eqref{zero distance bounded remainder}
gives $b_{\rm B}^L=\kappa\delta_{L,0}$ and $b_{\rm F}^L=0$, hence
\begin{align}
 \boxed{\ \kappa\,\delta_{L,0}+
 \sum_\D a_\D^{\rm B}\frac{2(s_\D)_L}{L!}\zeta(s_\D+L)=0\ },\qquad
 L\in2\mathbb Z_{\geq0},\\
 \boxed{\ \sum_\D a_\D^{\rm F}\frac{2(s_\D)_L}{L!}
 \operatorname{Li}_{s_\D+L}(-1)=0\ },\qquad
 L\in2\mathbb Z_{\geq0}.
 \label{bootstrap equation x=0}
\end{align}
Thus every positive-level zero-separation equation is source-free, and
only the bosonic level-zero equation contains a constant.  There is no
undetermined function in these equations, but this does not imply that each
spin-resolved $\mathcal B^{\widehat n,\widehat J}$ vanishes separately.
If the reconstruction assumptions fail, the general sourced equations
\eqref{zero distance sourced equations} must be used.
Appendix~\ref{app:ON-zero-distance} checks all levels in the critical
large-$N$ $O(N)$ model.  The resummed Lee--Yang example in
appendix~\ref{app:LY-zero-distance} instead describes the conformal-vacuum
cylinder state, not the physical thermal state.

\section{Analytical solutions of the thermal Polyakov bootstrap equations}
\label{sec:block-applications}
We apply the thermal Polyakov bootstrap equations
\eqref{new Polyakov equation} and \eqref{new Polyakov equation fermion}
to analytically tractable thermal CFTs.
The examples below revisit known thermal correlators and OPE data: the bosonic
benchmarks studied using thermal inversion and dispersion relations
\cite{Iliesiu:2018fao,Barrat:2025nvu}, and their fermionic counterparts
\cite{Petkou:2018ynm,David:2023uya}.  We reorganize these results as tests of the
thermal Polyakov cancellation equations and their source terms.
Throughout we set $\beta=1$, specify the external dimension and statistics,
and use the vacuum spectrum and OPE data as input.  The double-twist basis organizes
the equations by $L=2\widehat n+\widehat J$: zeta or polylogarithm zeros make some
sectors finite, while infinite sectors require a common regulator before any sums
are interchanged.  We include the source $\mathcal B^{\widehat n,\widehat J}$ unless
clustering and the interchange of sums justify its absence; in particular, low-spin
data need not follow from homogeneous equations alone.


\subsection{Generalized free scalar: solutions from a diagonal Polyakov matrix}
\label{sec:gff-block-example}

The generalized-free scalar is an exactly solvable reference point for
large-spin expansions, with its thermal correlator obtained by summing vacuum
images.  Its thermal OPE coefficients were recovered using inversion
\cite{Iliesiu:2018fao} and studied through image-based reconstruction
\cite{Barrat:2025nvu}; here they calibrate the normalization and the finite
cancellation condition.

Its spectrum is
$\mathds1+[\phi\phi]_{n,J}$ with
$\D_{n,J}=2\D_\f+2n+J$ and even $J$.  Inserting this spectrum into
\eqref{unphysical coefficient new} shows that the zeta argument of a bilinear in the
equation at level $L$ is
\begin{align}
 s_{\widehat n\widehat J}(\D_{n,J})=L-(2n+J).
 \label{GFF image exponent}
\end{align}
If $2n+J>L$, this is a negative even integer and the contribution vanishes because
$\zeta(-2q)=0$ for $q\geq1$.  Thus every equation is finite:
\begin{align}
 \mathcal A_{\mathds1}^{\widehat n,\widehat J}
 +\sum_{\substack{n\geq0,\ J\in2\bb Z_{\geq0}\\2n+J\leq L}}
 a_{2\D_\f+2n+J,J}\,
 \mathcal A_{2\D_\f+2n+J,J}^{\widehat n,\widehat J}=0.
 \label{free Polyakov equation new}
\end{align}
At $L=0$, the two nonzero kernels are
$\mathcal A_{\mathds1}^{0,0}=2\zeta(2\D_\f)$ and
$\mathcal A_{2\D_\f,0}^{0,0}=-1$, so
\begin{align}
 a_{2\D_\f,0}=2\zeta(2\D_\f).
 \label{GFF scalar coefficient}
\end{align}
By the all-level identity \eqref{double twist diagonal kernel}, the
classical-input matrix in the thermal-block basis is exactly $-I$; the triangular
recursion appears only after changing to the monomial basis.  The solution is
therefore the identity-generated coefficient,
\begin{align}
 a_{2\D_\f+2n+J,J}
 =2\zeta(2\D_\f+2n+J)
 \frac{(J+\nu)(\D_\f)_{J+n}(\D_\f-\nu)_n}
      {n!(\nu)_{J+n+1}},
 \qquad \nu=\frac{d-2}{2},
 \label{free thermal coefficient}
\end{align}
in agreement with the thermal Lorentzian inversion formula
\cite{Iliesiu:2018fao}.  The first two coefficients at the next level provide useful
implementation checks:
\begin{align}
 a_{2\D_\f+2,0}
 &=2\zeta(2\D_\f+2)\,
   \frac{\D_\f(\D_\f-\nu)}{\nu+1},\nn
 a_{2\D_\f+2,2}
 &=2\zeta(2\D_\f+2)\,
   \frac{\D_\f(\D_\f+1)}{\nu(\nu+1)}.
 \label{GFF low levels}
\end{align}
As a direct check in the monomial basis, at even degree $L=\a+\b$ the same
spectrum reduces \eqref{Polyakov bootstrap} to
\begin{align}
 \frac{(\D_\f)_\a(\D_\f)_\b}{\a!\b!}\,
 \zeta(L+2\D_\f)
 -\frac12\sum_{n=0}^{\min(\a,\b)}
 \frac{(\nu)_{\a-n}(\nu)_{\b-n}}
      {(\a-n)!(\b-n)!}\,
 a_{2\D_\f+L,L-2n}=0.
 \label{GFF bootstrap equation}
\end{align}
This finite system is triangular in spin and is satisfied by
\eqref{free thermal coefficient}, making the equivalence of the two formulations
explicit in this example.

At the scalar unitarity bound $\D_\f=\nu$, all trajectories with $n>0$ vanish and
only the $n=0$ current tower remains.  The massless scalar in $d=3$ is exceptional:
$\D_\f=\frac12$ makes the zero-mode coefficient $\zeta(1)$ divergent, reflecting the
infrared obstruction to the unregulated thermal image sum.  More generally, the image
sum is first defined in its convergence domain and then meromorphically continued.

Note that there is no contradiction when a generated block in
\eqref{double twist Polyakov decomposition} has the same dimension as a physical
generalized-free operator.  Comparing the coefficient of that block in
$g=\sum a_{\cal O}P_{\cal O}$ with the coefficient in the physical OPE cancels the
direct $a_{n,J}$ contribution on the two sides and leaves exactly
\eqref{free Polyakov equation new}.  Equivalently, one may work at generic dimensions
and take the coincident limit at the end.

\subsection{\texorpdfstring{The critical $O(N)$ model in $4-\epsilon$ dimensions}
{The critical O(N) model in 4-epsilon dimensions}}
\label{sec:ON-epsilon-block-example}

The Wilson--Fisher fixed point is a perturbatively accessible interacting
deformation of the free scalar theory near four dimensions.  We revisit the
thermal correlator and OPE trajectories studied using analytic bootstrap and
Feynman diagrams in Ref.~\cite{Barrat:2025nvu}.  This tests how anomalous
dimensions modify the classical cancellations, including the infrared-sensitive
scalar sector.

Let
\begin{align}
 c_N=\frac{N+2}{N+8},\qquad
 \D_\f=1-\frac{\epsilon}{2}+O(\epsilon^2).
 \label{WF coupling definitions}
\end{align}
To first order, the relevant singlet sectors in $\phi_i\times\phi_j$ are the
leading-twist bilinears $\mathcal O_{0,J}=[\phi\phi]_{0,J}$ and the next trajectory
$\mathcal O_{1,J}=[\phi\phi]_{1,J}$, with even $J$.  The latter is often denoted
schematically by $\partial^J\phi^4$; more precisely, it is classically represented by
$\phi\,\partial^J\Box\phi$ and is related to
$\epsilon\,\phi\,\partial^J\phi^3$ by the equation of motion and multiplet
recombination \cite{Rychkov:2015naa}.  The required spectral input is
\begin{align}
 \D_{0,0}&=2-\epsilon+c_N\epsilon+O(\epsilon^2),\nn
 \D_{0,J\geq2}&=2+J-\epsilon+O(\epsilon^2),\label{WF spectral input} \\
 \D_{1,J}&=4+J+O(\epsilon).
 \nonumber
\end{align}
The $O(\epsilon)$ shift of $\D_{1,J}$ is immaterial here because its thermal
coefficient already starts at $O(\epsilon)$.  Below, $a_{r,J}$ and
$\mathcal A_{r,J}$ abbreviate the data and kernel of $\mathcal O_{r,J}$.

For a target of level $L$, the zeroth-order zeta factors are respectively
$\zeta(L-J)$ and $\zeta(L-J-2)$.  Their trivial zeros imply
$J\leq L$ for $\mathcal O_{0,J}$ and $J\leq L-2$ for
$\mathcal O_{1,J}$ at this order in the regulated perturbative expansion.
Collecting the scalar input and any arc/reordering contribution,
\begin{align}
 \mathcal S^{\widehat n,\widehat J}
 \equiv\mathcal B^{\widehat n,\widehat J}
+a_{\phi^2}\mathcal A_{\phi^2}^{\widehat n,\widehat J},
 \label{WF source definition}
\end{align}
the finite equations are
\begin{align}
 \mathcal A_{\mathds1}^{\widehat n,\widehat J}
 +\mathcal S^{\widehat n,\widehat J}
 +\sum_{\substack{2\leq J\leq L\\J\ {\rm{even}}}}
 a_{0,J}\mathcal A_{0,J}^{\widehat n,\widehat J}
 +\sum_{\substack{0\leq J\leq L-2\\J\ {\rm{even}}}}
 a_{1,J}\mathcal A_{1,J}^{\widehat n,\widehat J}
 =O(\epsilon^{3/2}).
 \label{new Polyakov equation 4-epsilon}
\end{align}
For the trajectories below we assume
$\mathcal B^{\widehat n,\widehat J}=O(\epsilon^{3/2})$ at every nonconstant
level $L\geq2$, and use $a_{\phi^2}=2\zeta(2)+O(\sqrt\epsilon)$.
The scalar-input kernel at these levels starts at $O(\epsilon)$, so its
infrared-sensitive $O(\sqrt\epsilon)$ correction first enters at
$O(\epsilon^{3/2})$.  Expanding the finite kernels then gives
\begin{align}
 a_{0,J\geq2}
 &=2\zeta(J+2)+\epsilon\left[
 -2\zeta'(J+2)
 -\frac{\pi^2}{3}\,c_N\,\frac{\zeta(J)}{J}\right]
 +O(\epsilon^{3/2}),\nn
 a_{1,J\geq0}
 &=\epsilon\,\frac{\pi^2}{3}\,c_N\,
 \frac{\zeta(J+2)}{J+2}
 +O(\epsilon^{3/2}),
 \qquad J\ {\rm{even}}.
 \label{WF thermal trajectories}
\end{align}
They agree with the analytic thermal bootstrap and the direct finite-temperature
calculation \cite{Barrat:2025nvu,Helton:2024critical}.  At the first nontrivial level,
the interaction corrections
to $a_{0,2}$ and $a_{1,0}$ at order $O(\epsilon)$ are
\begin{align}
 a_{0,2}^{(1)}
 =-\frac{\pi^4}{36}c_N,\qquad
 a_{1,0}^{(1)}=\frac{\pi^4}{36}c_N,
 \label{WF first interaction shifts}
\end{align}
which is a compact check of the mixing between the two trajectories.

The scalar $\phi^2$ is different.  If one sets the compensating source $\mathcal B$ to
zero while including the $\phi^2$ term in the homogeneous spectral sum, the formal
solution is 
\begin{align}
 a_{\phi^2}|_{\mathcal B=0}
 =2\zeta(2)-2\zeta'(2)\epsilon
 +\frac{\pi^2\log(2\pi)}{3}c_N\epsilon+\cdots.
 \label{formal phi2 coefficient}
\end{align}
This is not a source-independent bootstrap prediction.  In the convention of
Ref.~\cite{Helton:2024critical}, resumming the thermal zero mode produces a leading
shift $-\pi\mth$.  Since $\mth=O(\sqrt\epsilon)$, that contribution is nonanalytic in
the naive $\epsilon$ expansion; assigning further error terms requires a specified
resummation scheme.  Thermal crossover calculations already exhibit the need to
treat these infrared contributions separately \cite{Sachdev:1996crossovers}.
The source assumption above is consistent with the resummed zero-mode term:
its leading $O(\sqrt\epsilon)$ contribution is constant, and its first
nonconstant Taylor coefficient is $O(\epsilon^{3/2})$.
Thus \eqref{WF thermal trajectories} is insensitive to this leading infrared
correction, but not to an arbitrary compensating source.  The scalar equation
must be supplemented by thermal infrared data.

\subsection{\texorpdfstring{The critical $O(N)$ model at infinite $N$}
{The critical O(N) model at infinite N}}
\label{sec:ON-largeN-block-example}

The three-dimensional $O(N)$ model is a standard setting for quantum-critical
thermal physics, including applications to antiferromagnets
\cite{Sachdev:1992py,Chubukov:1993critical}.  At infinite $N$, its thermal saddle
generates a self-consistent screening mass, and its correlator and higher-spin
one-point data are known \cite{Iliesiu:2018fao,Petkou:2018ynm,Barrat:2025nvu}.
We use these results to test the regulated cancellation hierarchy and recovery
of the known gap equation, first keeping $2<d<4$ general and then specializing
to $d=3$.

We set $\D_\f=(d-2)/2$ and $\beta=1$.
The auxiliary field has
$\D_\sigma=2$, so $\D_{\sigma^p}=2p$, and
$\langle\sigma\rangle_\beta=\mth^2$.  Large-$N$ factorization gives
\begin{align}
 a_{\sigma^p}
 =\frac{\mth^{2p}}{4^p p!(\frac{4-d}{2})_p}.
 \label{large N sigma coefficients}
\end{align}
This sector is infinite in every bootstrap equation and must be summed with a common
analytic regulator.

The origin of the compensating term can be seen directly in momentum space.  With
$\omega_n=2\pi n$,
\begin{align}
 a_{\sigma^p}\widetilde{\mathcal P}_{p}(\omega_n,k)
 =\mathcal N_d\frac{(-\mth^2)^p}
 {(k^2+\omega_n^2)^{p+1}},
 \qquad
 \mathcal N_d=\frac{4\pi^{d/2}}{\Gamma(\frac{d-2}{2})}.
 \label{large N momentum block}
\end{align}
Summing in $p$ before the Fourier transform produces the massive denominator
$1/(k^2+\omega_n^2+\mth^2)$, whereas transforming term by term and then summing
expands around the massless denominator.  In the dimensional/analytic prescription
used here, the required completion is supported at $\omega_0=0$ and equals
\begin{align}
 G_{\rm{red}}(z,\zb)
 =\frac{\sqrt{\pi}\,\Gamma(\frac{3-d}{2})\mth^{d-3}}
 {2^{d-3}\Gamma(\frac{d-2}{2})}\,
 {}_0F_1\left(;\frac{d-1}{2};
 \frac{\mth^2|\mathbf x|^2}{4}\right).
 \label{large N zero mode source}
\end{align}
Projecting this function onto the redundant double-twist blocks gives
\begin{align}
 \mathcal B^{\widehat n,\widehat J}
 ={}&(-1)^{\widehat J/2}
 \frac{\sqrt{\pi}\,\widehat J!(\widehat J+\frac{d-2}{2})
 \Gamma(\frac{3-d}{2})\Gamma(\frac{d-1}{2})}
 {2^{d-3+2\widehat n+2\widehat J}
 \widehat n!(\frac{\widehat J}{2})!
 \Gamma(\frac{d-1+\widehat J}{2})
 \Gamma(\frac d2+\widehat n+\widehat J)}
 \mth^{d-3+2\widehat n+\widehat J}.
 \label{large N block source}
\end{align}
The standard equation therefore reads
\begin{align}
 \mathcal A_{\mathds1}^{\widehat n,\widehat J}
 &+\sum_{\substack{n\geq0,\ J\ {\rm{even}}\\2n+J\leq L\\(n,J)\neq(0,0)}}
 a_{[\phi\phi]_{n,J}}\,
 \mathcal A_{[\phi\phi]_{n,J}}^{\widehat n,\widehat J}
 +\sum_{p=1}^{\infty}a_{\sigma^p}
 \mathcal A_{\sigma^p}^{\widehat n,\widehat J}
 +\mathcal B^{\widehat n,\widehat J}=0.
 \label{Polyakov equation ON large N new}
\end{align}
Only the bilinear sum is finite at fixed $L$.

There is no primary $[\phi\phi]_{0,0}$ in the critical spectrum.  Consequently, the
$(\widehat n,\widehat J)=(0,0)$ equation is a compatibility condition,
\begin{align}
 2\sum_{p=0}^{\infty}
 \frac{\zeta(d-2-2p)}
 {4^p p!(\frac{4-d}{2})_p}\mth^{2p}
 +\frac{\sqrt{\pi}\Gamma(\frac{3-d}{2})}
 {2^{d-3}\Gamma(\frac{d-2}{2})}\mth^{d-3}=0,
 \label{large N O gap zeta}
\end{align}
where $p=0$ represents the identity.  This is the thermal gap equation
\cite{Sachdev:1992py,Petkou:2018ynm}.  An equivalent
form, obtained directly from the massive image correlator, is
\begin{align}
 \frac{\Gamma(\frac{2-d}{2})}{2^{(d+2)/2}}
 +\sum_{q=1}^{\infty}
 \frac{K_{\frac{d-2}{2}}(q\mth)}
 {(q\mth)^{\frac{d-2}{2}}}=0.
 \label{large N O gap Bessel}
\end{align}
The Bessel image sum converges for positive $\mth$, whereas the zeta representation
is analytically continued as a complete expression.  In particular, at $d=3$
the pole in $\Gamma((3-d)/2)$ cancels the pole associated with $\zeta(d-2)$ only in
the complete equation.  The finite result is
\begin{align}
 2\operatorname{Li}_1(e^{-\mth})=\mth,\qquad
 \mth=2\log\left(\frac{1+\sqrt5}{2}\right).
 \label{large N O gap d3}
\end{align}

Once \eqref{large N O gap zeta} is imposed, the higher-level equations give, for
$(n,J)\neq(0,0)$,
\begin{align}
 a_{[\phi\phi]_{n,J}}
 ={}&\big(1+(-1)^J\big)
 \frac{2(J+\frac{d-2}{2})
 \mth^{\frac{d-2}{2}+J+2n}}
 {2^{\frac{d-2}{2}+J+2n}n!\,
 \Gamma(\frac d2+J+n)}
 \sum_{q=1}^{\infty}
 \frac{K_{J+\frac{d-2}{2}}(q\mth)}
 {q^{\frac{d-2}{2}}}
 \nn
 &-\delta_{J,0}
 \frac{\Gamma(\frac{4-d}{2})\mth^{d-2+2n}}
 {2^{d-2+2n}n!\Gamma(\frac d2+n)}.
 \label{large N O bilinear coefficients}
\end{align}
For $n>0$, this coefficient can represent a sum over degenerate
primaries rather than a separately resolved operator.

Several deductions are immediate in $d=3$.  Equation
\eqref{large N sigma coefficients} reduces to
$a_{\sigma^p}=\mth^{2p}/(2p)!$.  More explicitly,
\begin{align}
 a_{[\phi\phi]_{n,0}}
 =\frac{\mth^{2n}}{(2n+1)!}
 \left[2\operatorname{Li}_1(e^{-\mth})-\mth\right],
 \label{large N O d3 scalar coefficients}
\end{align}
so every $J=0$ coefficient vanishes when the gap equation is satisfied.  For $n=0$ and even
$J\geq2$, the answer becomes
\begin{align}
 a_J=\sum_{r=0}^{J}
 \frac{2^{r+1}}{r!}
 \frac{(J-r+1)_r}{(2J-r+1)_r}
 \mth^r\operatorname{Li}_{J+1-r}(e^{-\mth}).
 \label{large N O d3 currents}
\end{align}
For $J=2$, \eqref{large N O gap d3} reduces this to
$a_T=8\zeta(3)/5$, consistent with the thermal polylogarithm identities of
Ref.~\cite{Sachdev:1993polylog}, while the $\mth\to0$ limit of the general result recovers the
generalized-free current tower.  Thus the lowest consistency equation fixes the
thermal scale, and the remaining hierarchy fixes the accessible one-point data.
Appendix~\ref{app:ON-zero-distance} provides an independent reconstruction at zero spatial separation
and exhibits its subtraction constant explicitly.  Beyond leading order, the thermal
free energy and higher-spin one-point functions of large-$N$ vector models have been
computed in Ref.~\cite{Diatlyk:2023finiteN}; these offer further tests of the sourced
bootstrap hierarchy.

\subsection{Exact two-dimensional cylinder correlators}
\label{sec:two-dimensional-examples}

In a unitary two-dimensional CFT, the plane-to-cylinder conformal map fixes
the thermal two-point function of identical Virasoro primaries on the infinite
spatial line \cite{Iliesiu:2018fao}.  We revisit the $\D_\f=1,2$ examples
studied in Ref.~\cite{Barrat:2025nvu}, whose exact correlators test the
spin-resolved cancellation equations without a perturbative approximation.

Let $\phi$ have weights $(h,\bar h)$ with $h=\bar h=\D_\f/2$.  The exponential map
gives \cite{Belavin:1984cft}
\begin{align}
 g(z,\zb)=\pi^{2\D_\f}
 \csc^{\D_\f}(\pi z)\csc^{\D_\f}(\pi\zb).
 \label{2D 2pt}
\end{align}
For nonzero spin, the usual $d>2$ block normalization degenerates as
$\nu\to0$.  A finite convention, including a separate normalization for $J=0$, is
\begin{align}
 f_{\D,J}^{2{\rm D}}(z,\zb)
 &=\frac{(1+\delta_{J,0})J!}{2}
 \lim_{d\to2}\frac{f_{\D,J}(z,\zb)}{(\frac{d-2}{2})_J}\nn
 &=\frac12(z\zb)^{\frac\D2-\D_\f}
 \left[\left(\frac z\zb\right)^{J/2}
 +\left(\frac\zb z\right)^{J/2}\right].
 \label{2D thermal block}
\end{align}
We use the corresponding finite kernel
\begin{align}
 \big(\mathcal A_{\D,J}^{\widehat n,\widehat J}\big)^{2{\rm D}}
 =\frac{(1+\delta_{J,0})J!}
 {(1+\delta_{\widehat J,0})\widehat J!}
 \lim_{d\to2}
 \frac{(\frac{d-2}{2})_{\widehat J}}
 {(\frac{d-2}{2})_J}
 \mathcal A_{\D,J}^{\widehat n,\widehat J}.
 \label{2D Polyakov kernel}
\end{align}

The coefficient extraction is especially simple in chiral variables.  For even
$q\geq0$,
\begin{align}
 \pi\csc(\pi z)
 &=\sum_{q\in2\bb Z_{\geq0}}c_q^{(1)}z^{q-1},
 &c_q^{(1)}&=-2\operatorname{Li}_q(-1),\nn
 \pi^2\csc^2(\pi z)
 &=\sum_{q\in2\bb Z_{\geq0}}c_q^{(2)}z^{q-2},
 &c_q^{(2)}&=2(q-1)\zeta(q),
 \label{2D chiral expansions}
\end{align}
with $\operatorname{Li}_0(-1)=\zeta(0)=-\frac12$ understood by analytic
continuation.  Setting
$p=(\D+J)/2$ and $q=(\D-J)/2$, the half-symmetrized block gives
\begin{align}
 a_{\D,J}^{2{\rm D}}
 =\frac{2c_p c_q}{1+\delta_{J,0}}.
 \label{2D coefficient extraction}
\end{align}
Thus the support is
\begin{align}
 J\in2\bb Z_{\geq0},\qquad \D\geq J,\qquad
 \D-J=0\pmod4,
 \label{2D spectral support}
\end{align}
including the identity.  At fixed $(\D,J)$ these coefficients aggregate any
degeneracy in the vacuum module.  Our $J>0$ block is one half of the sum of the two
chiral monomials; this accounts for the factor of two relative to the convention used
in Ref.~\cite{Barrat:2025nvu}.

For $\D_\f=1$, equations \eqref{2D chiral expansions} and
\eqref{2D coefficient extraction} give
\begin{align}
 a_{\D,J}^{2{\rm D}}
 =\frac{8
 \operatorname{Li}_{\frac{\D-J}{2}}(-1)
 \operatorname{Li}_{\frac{\D+J}{2}}(-1)}
 {1+\delta_{J,0}},
 \label{2D coefficients dimension one}
\end{align}
on the support \eqref{2D spectral support}, and zero otherwise.  The first values are
\begin{align}
 a_{0,0}=1,\qquad a_{2,2}=\frac{\pi^2}{3},\qquad
 a_{4,0}=\frac{\pi^4}{36},\qquad
 a_{4,4}=\frac{7\pi^4}{180}.
 \label{2D dimension one low coefficients}
\end{align}
The lowest nontrivial cancellation is simply
$\pi^2/3-a_{2,2}=0$.

For $\D_\f=2$, one obtains
\begin{align}
 a_{\D,J}^{2{\rm D}}
 =\frac{2(\D-J-2)(\D+J-2)
 \zeta(\frac{\D-J}{2})\zeta(\frac{\D+J}{2})}
 {1+\delta_{J,0}},
 \label{2D coefficients dimension two}
\end{align}
again on \eqref{2D spectral support}.  Now
\begin{align}
 a_{0,0}=1,\qquad a_{2,2}=\frac{2\pi^2}{3},\qquad
 a_{4,0}=\frac{\pi^4}{9},\qquad
 a_{4,4}=\frac{2\pi^4}{15},
 \label{2D dimension two low coefficients}
\end{align}
and the $(\widehat n,\widehat J)=(0,0)$ equation reads
\begin{align}
 \frac{\pi^4}{45}+\frac{\pi^2}{3}a_{2,2}
 -a_{4,0}-a_{4,4}=0.
 \label{2D worked cancellation}
\end{align}
Direct substitution shows
\begin{align}
 \sum_{\{\D,J\}}a_{\D,J}^{2{\rm D}}
 \big(\mathcal A_{\D,J}^{\widehat n,\widehat J}\big)^{2{\rm D}}=0
 \, .
 \label{2D Polyakov bootstrap equation new}
\end{align}
for every $\widehat n\geq0$ and even $\widehat J\geq0$.  These exact correlators are
therefore useful residual tests, although a finite set of their equations need not
uniquely determine all coefficients at the same cutoff.

\subsection{Generalized free fermions}
\label{sec:fermionic-generalized-free}

Generalized free fermions provide an exact antiperiodic reference for thermal
one-point data \cite{David:2023uya}.  Following that work, we use the
derivative-contracted scalar structure $g_3$ to test the scalar Polyakov kernels;
this is not a treatment of the full spinor correlator.  The canonical
free-field limit of its normalization requires special care, as discussed below.
Let
$n_s$ be the spinor dimension and define
\begin{align}
 g_3(x)=\frac1{n_s}\operatorname{tr}
 \left\langle\partial_\mu\bar\psi(x)\gamma^\mu\psi(0)\right\rangle_\beta .
 \label{fermion derivative trace}
\end{align}
For a generalized-free spinor of dimension $\D_\ps$,
\begin{align}
 \widehat g_{\rm F}(x)
 \equiv\frac{g_3(x)}{d-1-2\D_\ps}
 =\sum_{q\in\bb Z}
 \frac{(-1)^q}
 {\big[(\tau+q)^2+|\mathbf x|^2\big]^{\D_\ps+\frac12}}.
 \label{normalized fermion scalar structure}
\end{align}
It obeys the antiperiodic KMS relation and is exactly a scalar image sum with
\begin{align}
 \D_{\rm F}=\D_\ps+\frac12,\qquad
 \D_{n,J}=2\D_\ps+1+2n+J,
 \qquad n\geq0,\quad J\in2\bb Z_{\geq0}.
 \label{fermion scalarized dimensions}
\end{align}
At level $L$, the polylogarithm in the bilinear kernel is
$\operatorname{Li}_{L-(2n+J)}(-1)$.  Since
$\operatorname{Li}_{-2q}(-1)=0$ for $q\geq1$, the recursion truncates to
$2n+J\leq L$:
\begin{align}
 \big(\mathcal A_{\mathds1}^{\widehat n,\widehat J}\big)_{\rm F}
 +\sum_{\substack{n\geq0,\ J\ {\rm{even}}\\2n+J\leq L}}
 a_{2\D_\ps+1+2n+J,J}
 \big(\mathcal A_{2\D_\ps+1+2n+J,J}^{\widehat n,\widehat J}\big)_{\rm F}=0.
 \label{fermion GFF block equation}
\end{align}
At $L=0$, the identity kernel is
$2\operatorname{Li}_{2\D_{\rm F}}(-1)$ and the scalar-bilinear kernel is $-1$.
The full solution is
\begin{align}
 a_{2\D_\ps+1+2n+J,J}
 ={}&2\operatorname{Li}_{2\D_\ps+1+2n+J}(-1)\nn
 &\times
 \frac{(J+\nu)(\D_\ps+\frac12)_{J+n}
       (\D_\ps-\frac{d-3}{2})_n}
      {n!(\nu)_{J+n+1}},
 \label{fermion GFF coefficients}
\end{align}
This follows from the bosonic solution \eqref{free thermal coefficient} by replacing
$\D_\f$ with $\D_{\rm F}=\D_\ps+\frac12$ and
$\zeta(s)$ with $\operatorname{Li}_s(-1)$, and agrees with
Ref.~\cite{David:2023uya}.

At the canonical free-field value $\D_\ps=(d-1)/2$, the unnormalized $g_3$ vanishes
by the Dirac equation.  Therefore \eqref{normalized fermion scalar structure} at that
point is an analytic-continuation, or null-descendant, normalization.  Its
coefficients encode a particular scalarized linear combination of fermionic one-point
data; they should not be described as the thermal coefficients of a nonzero,
standalone free-fermion scalar correlator.  Treating the nonvanishing vector structure
directly would require a spinning version of the present equations.

\subsection{\texorpdfstring{The large-$N$ Gross--Neveu fixed point in $d=5$}
{The large-N Gross--Neveu fixed point in d=5}}
\label{sec:GN-largeN-example}

The critical $U(N)$ Gross--Neveu model supplies an interacting fermionic test
with a thermal mass fixed by a saddle equation.  We revisit the scalar-trace
correlator and gap equation of Refs.~\cite{Petkou:2018ynm,David:2023uya} in
the prescribed large-$N$ continuation to $d=5$; no finite-$N$ ultraviolet
completion is assumed here.  This tests antiperiodic cancellation in the
presence of an infinite scalar sector.

We work at leading order in $1/N$, with $\D_\ps=\D_\f=2$.
Writing $n_s$ for the spinor dimension,
we normalize the scalar trace as
\begin{align}
 g_1(x)\equiv\frac1{n_s}\sum_{\alpha=1}^{n_s}
 \left\langle\bar\psi_\alpha(x)\psi_\alpha(0)\right\rangle_\beta.
 \label{GN normalized scalar trace}
\end{align}
The positive thermal mass will be denoted by $\mth$.  The scalar trace of the fermion
propagator is
\begin{align}
 g_1(z,\zb)
 =\frac{\mth^{d/2}}{(2\pi)^{d/2}}
 \sum_{q\in\bb Z}(-1)^q
 \frac{K_{\frac{d-2}{2}}\!
 \left(\mth\sqrt{(z+q)(\zb+q)}\right)}
 {\big[(z+q)(\zb+q)\big]^{\frac{d-2}{4}}}.
 \label{g1 large N}
\end{align}
Using $K_{3/2}(x)=\sqrt{\pi/(2x)}e^{-x}(1+1/x)$, this becomes in $d=5$
\begin{align}
 g_1(z,\zb)
 =\frac1{8\pi^2}\sum_{q\in\bb Z}(-1)^q e^{-\mth R_q}
 \left(\frac{\mth}{R_q^3}+\frac{\mth^2}{R_q^2}\right),
 \qquad R_q=\sqrt{(z+q)(\zb+q)}.
 \label{g1 d5 elementary}
\end{align}
The scalar saddle sector in this channel has $(\D,J)=(p,0)$ with odd $p\geq1$,
while the fermion bilinears have $(\D,J)=(4+2n+J,J)$ with $n\geq0$ and even
$J$.  The scalar trace contains no identity contribution.

The $q=0$ image gives scalar local coefficients
\begin{align}
 a_{p,0}^{(0)}
 =\frac{(-1)^p(p-2)\mth^p}{8\pi^2(p-1)!},
 \qquad p\geq1.
 \label{GN zero image coefficients}
\end{align}
These are intermediate coefficients of the zero image.  For even $p$ they combine
with the nonzero images into the even-dimensional bilinear coefficients below.
In particular,
\begin{align}
 a_{1,0}=\frac{\mth}{8\pi^2},\qquad a_{2,0}=0,\qquad
 a_{\D,0}=-\frac{\mth^\D}
 {8\pi^2(\D-1)(\D-3)!}\quad
 (\D\geq3\ {\rm{odd}}).
 \label{GN odd scalar coefficients}
\end{align}
Pairing the $q$ and $-q$ images removes odd Taylor degree from the nonzero-image
part.  The remaining support has even $J\geq0$ and
$\D-J\in2\bb Z_{\geq4}$, with total coefficients
\begin{align}
 a_{\D,J}
 ={}&\frac{\mth^{\D-J-4}}
 {2^{\D-J-4}
  (\frac{\D-J}{2}-2)!
  (J+\frac52)_{\frac{\D-J}{2}-2}}
 \nn
 &\times\left[
 \delta_{J,0}\frac{\mth^4}{24\pi^2}
 +\frac{\mth}{4\pi^2(2J+1)!!}
 \sum_{k=0}^{J+1}
 \frac{(2J+2-k)!\,\mth^k}
 {2^{J+1-k}(J+1-k)!k!}
 \operatorname{Li}_{J+3-k}(-e^{-\mth})\right].
 \label{GN even coefficients}
\end{align}
Here \emph{total} is essential: operators degenerate at fixed $(\D,J)$ are not
resolved by this correlator alone.  The coefficients not covered by
\eqref{GN odd scalar coefficients} or \eqref{GN even coefficients} vanish.

With the absent $(\D,J)=(4,0)$ bilinear omitted, the complete fermionic Polyakov
system for this correlator is therefore
\begin{align}
 0={}&\sum_{\substack{p\geq1\\p\ {\rm{odd}}}}a_{p,0}
 \big(\mathcal A_{p,0}^{\widehat n,\widehat J}\big)_{\rm F}
 +\sum_{\substack{n\geq0,\ J\in2\bb Z_{\geq0}\\(n,J)\neq(0,0)}}
 a_{4+2n+J,J}
 \big(\mathcal A_{4+2n+J,J}^{\widehat n,\widehat J}\big)_{\rm F},
 \nn[-2mm]
 &\hspace{6cm}\widehat n\geq0,
 \qquad\widehat J\in2\bb Z_{\geq0}.
 \label{GN full Polyakov equations}
\end{align}
The first sum is understood with the common analytic continuation inherited from the
symmetric image sum.  At fixed $(\widehat n,\widehat J)$ the second sum truncates.
There is neither an identity term nor an additional source
$\mathcal B_{\rm F}^{\widehat n,\widehat J}$.

The would-be bilinear $\mathcal O_0[0,0]$ at $(\D,J)=(4,0)$ is absent from the
critical spectrum.  The coefficient of this would-be block obtained from
\eqref{GN even coefficients} is
\begin{align}
 a_{4,0}
 &=\frac{\mth^4}{24\pi^2}
 +\frac{\mth}{4\pi^2}
 \left[\operatorname{Li}_3(-e^{-\mth})
 +\mth\operatorname{Li}_2(-e^{-\mth})\right]\nn
 &=\frac{\mth}{24\pi^2}
 \left[\mth^3+6\mth\operatorname{Li}_2(-e^{-\mth})
 +6\operatorname{Li}_3(-e^{-\mth})\right].
 \label{GN a40}
\end{align}
Its absence therefore imposes
\begin{align}
 \mth^3+6\mth\operatorname{Li}_2(-e^{-\mth})
 +6\operatorname{Li}_3(-e^{-\mth})=0,
 \qquad
 \mth=1.4805105937\ldots .
 \label{GN d5 gap equation}
\end{align}
This is precisely the large-$N$ thermal gap equation.  It also has a direct
Polyakov-bootstrap interpretation.  At $(\widehat n,\widehat J)=(0,0)$,
\begin{align}
 \big(\mathcal A_{\D,0}^{0,0}\big)_{\rm F}
 =2\operatorname{Li}_{4-\D}(-1).
 \label{GN lowest kernel}
\end{align}
The regulated sum of the odd-scalar contributions equals the expression
$a_{4,0}$ in \eqref{GN a40}; requiring the redundant coefficient to vanish is
therefore equivalent to \eqref{GN d5 gap equation}.

At fixed $(\widehat n,\widehat J)$, the even-dimensional bilinear sector truncates
because of the zeros $\operatorname{Li}_{-2q}(-1)=0$, whereas the odd scalar sector
remains infinite.  We sum the latter in increasing $\D$ using the analytic
continuation inherited from the symmetric image sum.  In $d=5$ its tail is
exponentially convergent since $\mth/\pi<1$.  At the root
\eqref{GN d5 gap equation}, truncating at $\D\leq20,30,40$ gives residuals of order
$10^{-10},10^{-13},10^{-16}$, respectively, for the equations
$(\widehat n,\widehat J)=(0,0),(0,2),(1,0),(0,4),(1,2),(2,0)$.  No additional source
$\mathcal B_{\rm F}$ is observed.  Structurally, this is consistent
with antiperiodic Matsubara frequencies, which have no zero mode analogous to
\eqref{large N zero mode source}.

Finally, define the normalized derivative trace $g_3$ as in
\eqref{fermion derivative trace}.  Away from contact terms, the massive Dirac
equation gives $g_3=\mth g_1$.  Its scalarization shifts
$(\D_\f,\D)\mapsto(\D_\f+\frac12,\D+1)$.  Both $2\D_\f-\D$ and
$(\D\pm J)/2-\D_\f$ are invariant under this shift, so the kernels in
\eqref{unphysical coefficient fermion new} are unchanged.  Hence the same Polyakov
equations hold for this second scalar structure.  In the free limit $\mth\to0$ it
vanishes, in agreement with the null-descendant caveat of
section~\ref{sec:fermionic-generalized-free}.

\section{Structures of the thermal Polyakov bootstrap equations and numerical prospects}
\label{sec:large-spin-numerics}
We study the analytical structure of the Polyakov matrix and its relation
to the large-spin expansions obtained from thermal Lorentzian inversion.
Two properties are particularly useful: its fixed-level large-spin
asymptotics and its reduction to $-I$ on the classical double-twist spectrum.
Together they suggest a numerical approach that uses known low-lying physical
data while eliminating a modeled large-spin tail.  We explain this
construction for the thermal 3D Ising model, separating the algebraic
reduction from the assumptions needed for a controlled numerical solution.

For this section, we denote a physical input operator by $(\Delta,\ell)$
and relabel the output double-twist indices $(\widehat n,\widehat J)$
as $(n,J)$.

\subsection{The Polyakov matrix at fixed radial level and large spin}
\label{sec:large-spin}

Write $(\Delta,\ell)$ for the input operator, $(n,J)$ for the output
double-twist label, and set
\begin{align*}
 p=\Delta_\phi-\frac{\Delta-\ell}{2},\qquad \nu=\frac{d-2}{2}>0.
\end{align*}
At fixed $n,\Delta,\ell$ and large even $J$, the compact kernel gives
\begin{align}
 \mathcal A_{\Delta,\ell}^{n,J}
 =2\zeta(J+2n+2\Delta_\phi-\Delta)
 \frac{\Gamma(\nu)}{\Gamma(p)}\frac{(\nu)_\ell}{\ell!}
 J^{p-\nu}
 \left[h_n+\frac{h_n b_n}{J}+O(J^{-2})\right],
 \label{large spin fixed level asymptotic} 
\end{align}
in which
\begin{align}
 h_n &=\frac{(p-\ell-\nu)_n}{n!},
 \nonumber\\
 b_n&=\frac{(p-\nu)(p+\nu-1)}2+n(p-\nu-1)     +\frac{\nu\ell(p+n-1)}{\nu+\ell-1}. \nonumber
\end{align}
The last term in $b_n$ is zero for $\ell=0$.  Thus every fixed-$n$ family
has the same leading spin power, with coefficient proportional to $h_n$,
when this coefficient is nonzero.  The derivation is given in
appendix~\ref{app:fixed-n-large-spin}. Exceptional parameters are taken in
the exact kernel before expanding.  This limit is not uniform for
$n\sim J$ or $\ell\sim J$.

For scalar input the finite sum can be evaluated exactly:
\begin{equation}
 \mathcal A_{\Delta,0}^{n,J}
 =2\zeta(J+2n+2\Delta_\phi-\Delta)
 \frac{(p)_J}{(\nu)_J}\frac{(p-\nu)_n}{n!}
 \frac{(J+p)_n}{(J+\nu+1)_n},
 \qquad p=\Delta_\phi-\frac{\Delta}{2}.
 \label{large spin scalar exchange exact}
\end{equation}
For example, the kinematic factors relative to $n=0$ are
$(p-\nu)(J+p)/(J+\nu+1)$ at $n=1$ and
$(p-\nu)_2(J+p)_2/[2(J+\nu+1)_2]$ at $n=2$.
The fermionic coefficients follow by replacing $\zeta(s)$ with
$\operatorname{Li}_s(-1)$.

\paragraph{Leading family and thermal inversion.}
At $n=0$, \eqref{double twist kinematic kernel} reduces to
\begin{align}
 \mathcal A_{\Delta,\ell}^{0,J}
 &=\frac{2\zeta(J+2\Delta_\phi-\Delta)}{(\nu)_J}
   \sum_{k=0}^{\ell}
   \frac{(\nu)_k(\nu)_{\ell-k}}{k!(\ell-k)!}(p-k)_J,
 &p&=\Delta_\phi-\frac{\Delta-\ell}{2}.
 \label{large spin leading family compact kernel}
\end{align}
As observed in Ref.~\cite{Barrat:2026jfg}, this is directly related to
thermal inversion.  For the full fixed-exchange formula (6.18) of
Ref.~\cite{Iliesiu:2018fao}, evaluated at the unshifted pole
$\bar h=J+\Delta_\phi$, with $d\bar h/dJ=1$ and the upper integration limit extended to infinity,
\begin{align*}
 4\pi K_J&=\frac{J!}{(\nu)_J},\qquad
 S^\infty_{-p+k,\Delta_\phi}(J+\Delta_\phi)=\frac{(p-k)_J}{J!},\\
 \mathcal A_{\Delta,\ell}^{0,J}
 &=\zeta(J+2\Delta_\phi-\Delta)\,\mathcal I_{\Delta,\ell}^{0,J}.
\end{align*}
Here $\mathcal I$ is the inversion contribution per unit input coefficient.
Since $\zeta(J+\mathrm{const.})=1+O(2^{-J})$, the two kernels have identical
algebraic $1/J$ expansions.  Formula (6.21) of that reference includes only
$k=0$; \eqref{large spin leading family compact kernel} includes the full
spin sum.

\paragraph{Ising light-operator seed.}
Set $\nu=1/2$, $\delta\equiv\Delta_\sigma=0.5181489$ and
$\Delta_\epsilon=1.412625$.  For $T$, $(\Delta,\ell)=(3,2)$,
\begin{align}
 \mathcal A_T^{0,J}
 &=2\zeta(J+2\delta-3)
   \frac{\frac38(p)_J+\frac14(p-1)_J+\frac38(p-2)_J}{(1/2)_J},
 &p&=\delta-\frac12.
 \label{large spin stress exchange exact}
\end{align}
The recurrence (2.32) of Ref.~\cite{Barrat:2026jfg} gives independently
\begin{align*}
 \mathcal A_T^{0,J}
 =\frac{2\zeta(J+2\delta-3)J!}{(1/2)_J}
 \left[
 \frac38\binom{\frac52-\delta}{J}
 +\frac14\binom{\frac32-\delta}{J}
 +\frac38\binom{\frac12-\delta}{J}
 \right],
\end{align*}
identical to the compact formula by $J!\binom{x}{J}=(-x)_J$ for even $J$.
Using
\begin{align}
 \frac{(q)_J}{(\nu)_J}
 =\frac{\Gamma(\nu)}{\Gamma(q)}J^{q-\nu}
 \left[1+\frac{(q-\nu)(q+\nu-1)}{2J}+O(J^{-2})\right].
 \label{large spin gamma ratio expansion}
\end{align}
the $n=0$ light-operator source is
\begin{align}
 s_J&\equiv
 \mathcal A_{\mathds1}^{0,J}
 +a_T\mathcal A_T^{0,J}
 +a_\epsilon\mathcal A_\epsilon^{0,J}
 \nn
 &=2\Bigg[
 J^{\delta-\frac12}
 \left(1.03545316+\frac{0.0001705301}{J}+O(J^{-2})\right)
 \nn
 &\hspace{8mm}
 +a_TJ^{\delta-1}
 \left(0.0121867943-\frac{0.0065623104}{J}+O(J^{-2})\right)
 \nn
 &\hspace{8mm}
 +a_\epsilon J^{\delta-\frac{\Delta_\epsilon}{2}-\frac12}
 \left(-0.289709710-\frac{0.0685987892}{J}+O(J^{-2})\right)
 \Bigg],
 \qquad J\ \text{even}.
 \label{large spin Ising asymptotic seed}
\end{align}
The exact formulas include the exponentially small image corrections.
All displayed terms agree with Eq.~(2.52) of Ref.~\cite{Barrat:2026jfg}
to its quoted precision except the sign of the subleading stress term:
\begin{align}
 c_T^{(1)}
 &=\frac{\sqrt{\pi}}{\Gamma(p)}
   \left[\frac{3}{16}\left(p-\frac12\right)^2
         +\frac{p-1}{4}\right],
 \qquad p=\delta-\frac12,
 \nn
 &=\underbrace{0.001414767892\ldots}_{k=0}
   \underbrace{{}-0.007977078268\ldots}_{k=1}
 =-0.006562310375\ldots.
 \label{large spin stress subleading sign}
\end{align}
The $k=1$ contribution exceeds the positive $k=0$ correction; $k=2$
starts at relative order $J^{-2}$.  Both the compact formula and the
recurrence therefore give the negative sign.

The positive $0.001414/J$ in Eq.~(6.78) of Ref.~\cite{Iliesiu:2018fao}
is precisely the $k=0$ term included in its leading-lightcone formula
(6.21).  Restoring $k=1$ in the full formula (6.18) gives
\eqref{large spin stress subleading sign}; the image factor and extension
of the upper integration limit affect only exponentially small terms in this fixed-input limit.

\Needspace{8\baselineskip}
\subsection{Towards solving the thermal Polyakov bootstrap equations numerically for the 3D Ising model}
\label{sec:ising-numerics}

The thermal 3D Ising model illustrates a possible numerical application
of the thermal Polyakov equations to an interacting CFT.  The rows of the
Polyakov matrix $\mathcal A_{\Delta,\ell}^{n,J}$ are labeled by
$\lambda=(n,J)$, whereas its columns are labeled by the physical operators
$\mathcal O=(\Delta,\ell)$ in the $\sigma\times\sigma$ OPE.
As an initial approximation, we separate the columns into the light sector
$\mathcal L=\{\mathds1,T,\epsilon\}$ and a remaining sector modeled by
double-twist trajectories.  The proposed solution has two stages: solve
the double-twist coefficients in terms of $a_T,a_\epsilon$; then substitute
them into the small-$(n,J)$ equations to constrain these two coefficients.
Throughout, $a_{\mathds1}=1$ and the vacuum spectrum is an input, not an
unknown of the thermal linear system.  Available Ising inputs include the
low-lying spectrum of Ref.~\cite{SimmonsDuffin:2016spectrum} and the improved
dimensions and OPE coefficients from mixed $\sigma,\epsilon,T_{\mu\nu}$
bootstrap \cite{Chang:2024IsingStressTensor}.

\paragraph{Splitting the matrix.}
Denote the modeled operators by $\mathcal O_\mu$, with
$\mu=(m,\ell)$, and their thermal coefficients by $y_\mu$.  Their dimensions
and Polyakov columns are
\begin{align}
 \Delta_\mu&=2\Delta_\sigma+2m+\ell+\gamma_\mu,
 &D_{\lambda\mu}
 &\equiv\mathcal A_{\Delta_\mu,\ell}^{n,J}
 =-\delta_{\lambda\mu}+R_{\lambda\mu}.
 \label{interacting double twist tail matrix}
\end{align}
The classical part follows from \eqref{double twist diagonal kernel}.
The correction $R$ is obtained from the exact kernel; for small
$\gamma_\mu$ it begins as
\begin{align}
 R_{\lambda\mu}
 =\gamma_\mu
 \left.\partial_\Delta\mathcal A_{\Delta,\ell}^{n,J}
 \right|_{\Delta=2\Delta_\sigma+2m+\ell}
 +O(\gamma_\mu^2).
 \label{numerical anomalous column correction}
\end{align}
Thus $D=-I$ on the classical double-twist lattice, while the anomalous
dimensions generate both diagonal and off-diagonal corrections.

Let $H$ be the set of modeled
double-twist labels and use the same labels for the rows that determine
$\mathbf y$.  Reserve a disjoint set $S$ of small-$(n,J)$ rows for the
light coefficients.  In the minimal two-parameter ansatz one can take
$S=\{(0,0),(0,2)\}$, excluding these two labels from $H$; $T$ and
$\epsilon$ are included explicitly, not counted again in the modeled
sector.  Consequently, the double-twist columns contribute
$D_{SH}=R_{SH}$ to the low rows and $D_{HH}=-I+R_{HH}$ to the matched rows,
as shown in figure~\ref{fig:polyakov-matrix-regions}.

\begin{figure}[htbp]
    \centering
\begin{tikzpicture}[x=1cm,y=1cm,font=\small]
  \fill[blue!5] (0,0) rectangle (6,2.5);
  \fill[blue!10] (0,1.25) rectangle (6,2.5);
  \fill[orange!10] (6,1.25) rectangle (10.2,2.5);
  \fill[teal!13] (6,0) rectangle (10.2,1.25);

  \draw[black!30,line width=.4pt] (6,0) -- (6,2.5);
  \draw[black!30,line width=.4pt] (0,1.25) -- (10.2,1.25);
  \draw[black!65,line width=.65pt]
    (.08,2.58) -- (-.13,2.58) -- (-.13,-.08) -- (.08,-.08);
  \draw[black!65,line width=.65pt]
    (10.12,2.58) -- (10.33,2.58) -- (10.33,-.08) -- (10.12,-.08);

  \node[align=center,text=blue!55!black] at (3,3.7)
    {Light operators $\mathcal L$};
  \node[align=center,text=teal!60!black] at (8.1,3.7)
    {Modeled double-twist operators};
  \node at (1,3.08) {$\mathds1$};
  \node at (3,3.08) {$T$};
  \node at (5,3.08) {$\epsilon$};
  \node at (8.1,3.08) {$\mathcal O_\mu,\quad\mu\in H$};

  \node[anchor=east,align=right] at (-.35,1.875)
    {$S$\\[2pt]\footnotesize low rows};
  \node[anchor=east,align=right] at (-.35,.625)
    {$H$\\[2pt]\footnotesize matched rows};

  \node at (1,1.875) {$\boldsymbol{\mathcal A}_{\mathds1}^{S}$};
  \node at (3,1.875) {$\boldsymbol{\mathcal A}_{T}^{S}$};
  \node at (5,1.875) {$\boldsymbol{\mathcal A}_{\epsilon}^{S}$};
  \node[text=orange!50!black] at (8.1,1.875) {$R_{SH}$};
  \node at (1,.625) {$\boldsymbol{\mathcal A}_{\mathds1}^{H}$};
  \node at (3,.625) {$\boldsymbol{\mathcal A}_{T}^{H}$};
  \node at (5,.625) {$\boldsymbol{\mathcal A}_{\epsilon}^{H}$};
  \node[text=teal!55!black] at (8.1,.625) {$-I+R_{HH}$};

  \node[anchor=north,text=black!75] at (1,-.4) {$1$};
  \node[anchor=north,text=black!75] at (3,-.4) {$a_T$};
  \node[anchor=north,text=black!75] at (5,-.4) {$a_\epsilon$};
  \node[anchor=north,text=black!75] at (8.1,-.4) {$\mathbf y(a_T,a_\epsilon)$};
  \node[anchor=east,font=\footnotesize,text=black!65] at (-.35,-.65)
    {coefficients};
\end{tikzpicture}
    \caption{Two-sector organization of the regulated Polyakov matrix.
    Columns are physical OPE operators; rows are double-twist labels
    $(n,J)$.  The matched rows $H$ determine $\mathbf y(a_T,a_\epsilon)$.
    Substituting this solution through the off-diagonal block $R_{SH}$
    gives the effective equations in the low rows $S$.  The source
    $\mathcal B$ is not a matrix column.}
    \label{fig:polyakov-matrix-regions}
\end{figure}

\paragraph{Solving the double-twist coefficients.}
Collect the light-operator contribution and the prescribed source into
\begin{align}
 q_\lambda(a_T,a_\epsilon)
 &\equiv\mathcal A_{\mathds1}^{\lambda}
 +a_T\mathcal A_T^{\lambda}
 +a_\epsilon\mathcal A_\epsilon^{\lambda}
 +\mathcal B^{\lambda}.
 \label{numerical light source}
\end{align}
The two groups of Polyakov equations are then
\begin{align}
 \mathbf q_S+R_{SH}\mathbf y&=0,
 &\mathbf q_H+(-I+R_{HH})\mathbf y&=0. 
 \label{numerical block system}
\end{align}
If the regulated matrix $I-R_{HH}$ is invertible, the second group gives
\begin{align}
 \mathbf y(a_T,a_\epsilon)
 &=(I-R_{HH})^{-1}
 \big(\boldsymbol{\mathcal A}_{\mathds1}^{H}
 +a_T\boldsymbol{\mathcal A}_T^{H}
 +a_\epsilon\boldsymbol{\mathcal A}_\epsilon^{H}
 +\boldsymbol{\mathcal B}^{H}\big).
 \label{interacting tail solution}
\end{align}
Every modeled double-twist thermal coefficient is therefore an affine
function of $a_T,a_\epsilon$, rather than an independent fitting parameter.
For a leading-family row $(0,J)\in H$, the same equation reads
\begin{align}
 y_{0,J}=s_J+\mathcal B^{0,J}
 +\sum_{\mu\in H}R_{(0,J),\mu}y_\mu,
 \label{large spin physical family equation}
\end{align}
where $s_J$ is the light-operator seed in
\eqref{large spin Ising asymptotic seed}.  This identifies the role of
subsection~\ref{sec:large-spin}: the light columns supply the large-spin
seed, while the double-twist subsystem incorporates the corrections from
the other operators in the modeled families.  At $R=0$ the solution
reduces to $y_{0,J}=s_J+\mathcal B^{0,J}$.

\paragraph{Substitution into the low rows.}
The first group in \eqref{numerical block system} is not solved until
the double-twist contribution $R_{SH}\mathbf y$ is included.  Substituting
\eqref{interacting tail solution} gives
\begin{align}
 \mathbf q_S(a_T,a_\epsilon)
 +R_{SH}(I-R_{HH})^{-1}\mathbf q_H(a_T,a_\epsilon)=0.
 \label{thermal Polyakov Schur system}
\end{align}
In particular, define the effective low-row columns and source by
\begin{align}
 \widetilde{\boldsymbol{\mathcal A}}_{\mathcal O}^{S}
 &=\boldsymbol{\mathcal A}_{\mathcal O}^{S}
 +R_{SH}(I-R_{HH})^{-1}\boldsymbol{\mathcal A}_{\mathcal O}^{H},
 \qquad \mathcal O\in\{\mathds1,T,\epsilon\},
 \nn
 \widetilde{\boldsymbol{\mathcal B}}^{S}
 &=\boldsymbol{\mathcal B}^{S}
 +R_{SH}(I-R_{HH})^{-1}\boldsymbol{\mathcal B}^{H}.
 \label{numerical effective low columns}
\end{align}
The remaining equations involve only the two unknown thermal coefficients:
\begin{align}
 a_T\widetilde{\boldsymbol{\mathcal A}}_T^{S}
 +a_\epsilon\widetilde{\boldsymbol{\mathcal A}}_\epsilon^{S}
 =-\widetilde{\boldsymbol{\mathcal A}}_{\mathds1}^{S}
  -\widetilde{\boldsymbol{\mathcal B}}^{S}.
 \label{numerical two coefficient system}
\end{align}
For $S=\{(0,0),(0,2)\}$ this is a $2\times2$ linear system.  If its
coefficient matrix is nonsingular, it determines $a_T,a_\epsilon$;
\eqref{interacting tail solution} then determines the modeled
double-twist coefficients.  The term
$R_{SH}(I-R_{HH})^{-1}\boldsymbol{\mathcal A}_{\mathcal O}^{H}$ is precisely
the contribution returned by the solved double-twist sector to each light
column.  It vanishes on the exact classical lattice, but contributes
for the interacting spectrum.  Thus the large-spin reconstruction and
the low-row constraints are coupled parts of the same Polyakov system.

\paragraph{Comparison with inversion and KMS.}
This construction can be compared directly with the thermal inversion
and KMS approach of Ref.~\cite{Iliesiu:2018zlz}.
The shared starting point is to express double-twist thermal data in
terms of light coefficients.  In Ref.~\cite{Iliesiu:2018zlz}, the thermal
Lorentzian inversion formula determines the family coefficients, which
are then inserted into the two-point correlator.  KMS covariance at
discrete points fixes the remaining parameters.
More precisely, stress-tensor consistency first relates $a_T$ to
$a_\epsilon$, leaving one parameter for the sampled-KMS fit.
Here the matched rows $H$ supply the solution
\eqref{interacting tail solution}, and substituting it into the low rows
$S$ gives \eqref{numerical two coefficient system}, without reconstructing
the correlator for a separate KMS fit.  Both methods allow thermal
coefficients of either sign.  The advantage of the Polyakov organization
is that reconstruction, off-diagonal feedback, and the remaining
constraints use the same Polyakov matrix.  It is a compact alternative,
with no separate choice of KMS sampling points.  Heavy-spectrum models
used to close KMS equations
\cite{Barrat:2025wbi,Barrat:2025nvu} provide complementary comparisons.

~

The two-sector split remains a spectral approximation: small double-twist
shifts are controlled at fixed radial level and large spin, and further
low-lying operators may need to be included explicitly.  The full family
sums and their feedback may require a common regulator consistent with
$\mathcal B$ and the prescriptions in appendix~\ref{sec:analytic-prescriptions}.  Convergence
of this regulated construction and stability of the extracted light
coefficients remain open; improved numerical precision is not established
by the algebraic reduction alone.  These questions are part of the
ongoing Ising study
\cite{Guo:2026IsingThermalPolyakov}.

\section{Summary and outlook}
\label{sec:outlook}

Our aim is to determine thermal one-point data from a prescribed vacuum
spectrum and the consistency of thermal two-point functions, without
assuming a definite sign for the thermal OPE coefficients.
The main results are the bosonic and fermionic thermal Polyakov bootstrap
equations \eqref{summary bosonic Polyakov equations} and
\eqref{summary fermionic Polyakov equations}.  Dispersion and the method
of images build KMS covariance into each block; matching their sum to the
physical OPE then requires cancellation of the generated classical
double-twist contributions.  This gives a compact linear system with
explicit matrix elements, including any required compensating terms.
We derive equivalent double-twist and monomial formulations and find
consistent solutions in generalized-free theories, critical vector
models, and two-dimensional CFTs.

In the earlier thermal bootstrap approach
\cite{Iliesiu:2018fao,Iliesiu:2018zlz}, thermal Lorentzian inversion
reconstructs large-spin families in terms of light thermal data.
The remaining coefficients are constrained by KMS crossing of the
reconstructed correlator, imposed at sampled points in the Ising
application.  More recent work developed dispersive image completion
and established the relation between individual thermal Polyakov blocks
and heavy thermal data
\cite{Barrat:2025nvu,Barrat:2026jfg}.  Building on these constructions, we
project the excess contribution of the full Polyakov sum onto each
classical double-twist block.  Reconstruction and the remaining OPE
constraints are thereby expressed as rows of the same explicit matrix,
rather than as a family reconstruction followed by a separate KMS fit.
This reorganizes thermal consistency without invoking positivity, which
is also absent in the earlier thermal approaches.  The analogy with the
vacuum Polyakov bootstrap is direct: KMS covariance of a two-point
function replaces crossing symmetry of a four-point function, while
spurious-term cancellation enforces the physical OPE in both cases
\cite{Polyakov:1974,Gopakumar:2016wkt,Mazac:2018qmi}.

We have identified properties of the Polyakov matrix that are useful
for analytical and numerical work.  The
light-operator columns supply the fixed-level large-spin seed, whose
leading-family expansion is related explicitly to thermal inversion.
On the classical double-twist spectrum the relevant submatrix is $-I$,
so the associated thermal coefficients can be eliminated directly.  For
interacting trajectories this becomes $-I+R$.  In our initial proposal
for the thermal 3D Ising CFT, the matched rows express
the modeled family coefficients in terms of $a_T,a_\epsilon$, and their
substitution through $R_{SH}$ gives the effective low-row equations.  The feedback term makes explicit
how the double-twist spectrum affects the light thermal data.  This suggests a
two-stage Ising calculation within one linear system; regularization of
the full family sums, control of omitted operators, and numerical stability
remain to be established before claiming a controlled solution.

Several questions follow naturally from this work:
\begin{itemize}
    \item \emph{A controlled thermal 3D Ising CFT solution.}
    The immediate aim is to turn precise vacuum Ising data into thermal
    one-point functions with assessed uncertainties.  The effective
    low-row system isolates the dependence on the modeled families,
    allowing their contribution to be varied and tested without fitting
    each thermal coefficient independently.  A joint $\sigma\sigma$ and
    $\epsilon\epsilon$ analysis can further constrain their shared
    one-point data.  We are pursuing this application in
    Ref.~\cite{Guo:2026IsingThermalPolyakov}; comparison with
    Ref.~\cite{Iliesiu:2018zlz} should use matched spectral inputs and
    include regulator, truncation, and conditioning uncertainties.


    \item \emph{Further analytic studies of thermal CFTs.}
    We have shown that the leading terms in the $4-\epsilon$ and large-$N$
    expansions are consistent with the thermal Polyakov bootstrap equations.
    Higher-order calculations in $\epsilon$ and $1/N$ can test the
    reconstruction prescription beyond these leading solvable examples.  Expanding the
    Polyakov matrix about its classical $-I$ sector organizes anomalous
    dimensions and thermal coefficients into successive linear problems.
    Operator mixing and infrared terms must be included at the same
    order, so comparison with existing subleading-$1/N$ results
    \cite{Diatlyk:2023finiteN} can also identify the required source terms.

    \item \emph{Momentum space and Mellin representations.}
    A representation adapted to frequency and momentum would connect
    the Euclidean constraints more directly to thermal response.
    Relating the cancellation equations to momentum-space thermal blocks
    \cite{Manenti:2019thermal} and OPE--quasinormal-mode sum rules
    \cite{Arnaudo:2026OPEQNM} could expose which spectral moments are fixed
    by spurious-term cancellation.  A thermal Mellin formulation may
    instead organize these constraints by poles and residues, making
    subtraction freedom explicit as in vacuum
    dispersive constructions \cite{Gopakumar:2021crossingMellin}.

    \item \emph{Spinning and charged thermal conformal correlators.}
    Currents and stress tensors give access to charge and energy
    response, motivating tensor-valued Polyakov blocks constrained by
    Ward identities.  Charged KMS twists
    \cite{David:2024chemical} would similarly modify the image weights.
    The resulting matrix equations could couple the allowed tensor
    structures and constrain their thermal data without positivity
    assumptions.  Stress-tensor correlators and holographic lightcone
    limits provide benchmarks
    \cite{Karlsson:2022thermalTT,Huang:2023ANEC,Esper:2023thermalTT}.
    Applications to transport require controlled Lorentzian continuation
    and the low-frequency limit
    \cite{CaronHuot:2009thermal,WitczakKrempa:2015dynamics}.

    \item \emph{Thermal Polyakov bootstrap study of holographic CFTs.}
    Holographic correlators supply independent thermal data and a way
    to interpret reconstruction ambiguities in the bulk
    \cite{Barrat:Holography2025}.  The cancellation equations can test
    whether multi-stress contributions and the associated double-twist
    sector reproduce the same low-row constraints as bulk calculations
    \cite{Niarchos:DoubleTwist2026}.  Spin-resolved data
    \cite{Buric:2026spinresolved} are especially useful for separating
    information fixed by the matrix from zero-mode freedom fixed by
    bulk regularity conditions.

    \item \emph{Defects and boundaries in thermal CFTs.}
    Thermal defects introduce local observables constrained by both KMS
    covariance and bulk-to-defect OPE consistency
    \cite{Barrat:2024defects}.  Combining image completion with the
    relevant OPE channels could turn these simultaneous requirements
    into coupled cancellation equations for bulk and defect thermal
    data.  Vacuum and boundary Polyakov methods
    \cite{Kangshabanik:2026truncated,Kangsabanik:2026bcft} suggest numerical
    strategies, although the thermal tensor structures and reconstruction
    terms must first be derived for this setting.
\end{itemize}

\Needspace{12\baselineskip}
\section*{Acknowledgements}
\addcontentsline{toc}{section}{Acknowledgements}
We thank Liangdong Hu and Wei Zhu for helpful discussions.
Some of the key results of this work were presented at the
\emph{2026 Mini-Symposium on Frontiers of Quantum Many-Body Physics} in February 2026.
ZL thanks the organizers for providing a stimulating environment.
ZL also thanks the Laboratoire de Physique de l'\'Ecole normale sup\'erieure
(LPENS), Paris, for its hospitality and support during the final stage of this work.
This work was supported by Southeast University startup funding
No.~4007022314 and by the National Natural Science Foundation of China
under Grant No.~12375061.

OpenAI Codex assisted, under author direction, with the calculations,
manuscript organization, and language.  The authors verified the text and
take full responsibility for the manuscript.

\appendix
\section{Analytic continuation and reconstruction prescriptions}
\label{sec:analytic-prescriptions}

This appendix collects the contour, subtraction and summation prescriptions used in
section~\ref{Derivation of the thermal Polyakov bootstrap equations}.  They specify the compensating terms appearing in the bootstrap
equations.

\subsection{Singular-point subtractions and continued crossed blocks}
\label{sec:endpoint-prescription}

The remainder $g_{\rm{hom}}$ in \eqref{dispersion plus arc} contains
integrals around branch points or poles, together with subtraction terms.  The
integration limits $w=0,r$ correspond to the origin and the branch
point at $r$; inversion also relates the origin to infinity.  When the
integrand is meromorphic near $w=0$ and infinity, the symmetry-folded
boundary term in the unsubtracted formula is
\begin{align*}
 R_{\rm{fold}}[g]=\Res_{w=0}\big[\mathcal K(z,\zb;w)\mathfrak g_r(w)\big]
 =\Res_{w=0}[K_0\mathfrak g_r]+\Res_{w=\infty}[K_0\mathfrak g_r].
\end{align*}
Indeed, $\mathcal K(w)=K_0(w)-w^{-2}K_0(1/w)$ and
$\mathfrak g_r(w)=\mathfrak g_r(1/w)$.  Thus this folded term already
includes the inversion-related infinity contribution; the same arc contribution must
not be added again.
For $g=1$ this is $1$, not zero.  At a branch point an ordinary residue is
inappropriate: the regulated small-contour integral must instead be evaluated together
with the local subtraction at that branch point.  Since $K_0=O(w^{-3})$ at infinity,
$\mathfrak g_r(w)=O(|w|^{2-\delta})$, $\delta>0$, uniformly on the large
arcs is a sufficient condition to drop those arcs; faster growth requires
subtractions or an explicit remainder.  Bounded-spin support of a subtraction
sector likewise requires an appropriate angular growth bound, not KMS alone
\cite{Alday:2020eua}.

Analytic subtraction defines the continued integral without discarding
its divergent local terms.  For example, if $F$ is regular at $t=0$,
with $F_j=F^{(j)}(0)/j!$, and the integrand is integrable at $t=1$,
one continues
\begin{align*}
 \int_0^1t^{u-1}F(t)\,{\rm d}t
 =\int_0^1t^{u-1}\left[F(t)-\sum_{j=0}^{N-1}F_jt^j\right]{\rm d}t
 +\sum_{j=0}^{N-1}\frac{F_j}{u+j},\qquad \Re u+N>0.
\end{align*}
If both integration limits are singular, split the interval and apply
the subtraction locally near each limit.  Logarithmic terms follow by differentiating the exponents.
At parameter poles, all regulated sectors and counterterms are combined before
taking a finite part.  This specifies the local continuation; it does not
establish convergence of an infinite spectral sum.

For the crossed block in \eqref{continued crossed block}, put
$a=(\D-J)/2-\D_\f$ and $b=a+J$.  The generic integrand behaves as
$w^{-b-1}$ at zero and $(r-w)^a$ at $r$.  Ordinary convergence requires
$\Re b<0$ and $\Re a>-1$, with no common strip for $J\geq1$.
We therefore expand the finite Gegenbauer polynomial, pair exchanged
monomials, continue their two exponents independently from
$-1<\Re p,\Re q<0$, and only then impose their physical values.
The local subtractions above implement this continuation.
For a symmetrized monomial pair the finite contour deformation first gives
\eqref{continued two image identity} in the common strip, and the identity
then continues analytically.

For polynomial inputs this prescription includes small-contour and
subtraction contributions that the pointwise discontinuity loses: for $f=1$, the continued two-image
expression is $2$ although the discontinuity is zero.  This defines the
individual continued block; it does not assert that the full correlator's
origin residue vanishes.

\subsection{Interchanging spectral and image sums}
\label{sec:summation-prescription}

Vanishing of a projected reordering defect requires control before the
regulator is removed.  For example, write $c_{\mathcal O,m}^{\a,\b}(\epsilon)$
for the regulated coefficient from one nonzero image.  After explicit pole
counterterms have been combined, a sufficient condition is a summable majorant
\begin{align*}
 \sum_{\mathcal O,m\neq0}\sup_{|\epsilon|<\epsilon_0}
 \left|a_{\mathcal O}(\epsilon)
 c_{\mathcal O,m}^{\a,\b}(\epsilon)\right|<\infty,
\end{align*}
together with dominated contour limits and locally uniform convergence allowing
the coefficient extraction.  Fubini's theorem and dominated convergence then give
\begin{align}
    [z^\a\zb^\b]G_{\rm{reorder}}=0
    \,,\label{finiteness condition}
\end{align}
while the homogeneous contribution can remain.  Finiteness of the set of
nonzero coefficients only \emph{after} analytic continuation, for example from
zeta zeros, is not by itself such a bound.  Polynomial benchmarks can instead
be checked directly in their specified image prescription.  In an interacting
theory the existence and value of $G_{\rm{red}}$ must not be inferred from those benchmarks.

\section{Kernel derivations}
\label{app:kernel-projections}

\subsection{Monomial coefficients}
\label{app:monomial-kernel}

The Gegenbauer expansion gives
\begin{align}
 f_{\D,J}(z,\zb)
 =\sum_{k=0}^{J}
 \frac{(\nu)_k(\nu)_{J-k}}{k!(J-k)!}
 z^{\frac{\D-J}{2}-\D_\f+k}
 \zb^{\frac{\D+J}{2}-\D_\f-k}
 \,.
 \label{thermal block finite expansion}
\end{align}
Pairing the positive and negative images,
\begin{align}
P_{\D,J}^{\rm X}(z,\zb)
=f_{\D,J}(z,\zb)
+\sum_{m=1}^{\infty}\eta_X^m
\left[f_{\D,J}(z+m,\zb+m)+f_{\D,J}(z-m,\zb-m)\right],
\label{paired ordinary images}
\end{align}
where $\eta_X$ is the statistics sign in \eqref{thermal covariance unified}, and
$P^{\rm B}=P$.  The generalized binomial theorem and
\begin{align}
\sum_{m=1}^{\infty}\frac1{m^s}=\zeta(s),\qquad
\sum_{m=1}^{\infty}\frac{(-1)^m}{m^s}
=\operatorname{Li}_s(-1),
\label{image Dirichlet series}
\end{align}
give the finite sum
\begin{align}
\big(A_{\D,J}^{\a,\b}\big)_{\rm X}
={}&\big(1+(-1)^{\a+\b}\big)
\mathscr Z_{\rm X}\!\left(s_{\a\b}(\D)\right)
\sum_{k=0}^{J}\frac{(\nu)_k(\nu)_{J-k}}{k!(J-k)!}
\nn
&\times
\binom{\frac{\D+J}{2}-\D_\f-k}{\a}
\binom{\frac{\D-J}{2}-\D_\f+k}{\b},
\qquad {\rm X}={\rm B},{\rm F},
\label{monomial finite sum kernel}
\end{align}
Here $\mathscr Z_{\rm B}(s)=\zeta(s)$,
$\mathscr Z_{\rm F}(s)=\operatorname{Li}_s(-1)$ and $s_{\a\b}$ is defined
in \eqref{monomial image exponent}.  The sums first converge for
$\operatorname{Re}s>1$; elsewhere the full expression is continued.

Writing $u=\D_\f-(\D+J)/2$, $v=\D_\f-(\D-J)/2$, the identities
\begin{align*}
 (u+k)_\a=(u)_\a\frac{(u+\a)_k}{(u)_k},\qquad
 (v-k)_\b=(v)_\b\frac{(1-v)_k}{(1-v-\b)_k}
\end{align*}
turn the $k$-sum into
\begin{align}
H_{\D,J}^{\a,\b}
={}&\frac{1}{\a!\b!}\frac{(\nu)_J}{J!}
\left(\D_\f-\frac{\D+J}{2}\right)_\a
\left(\D_\f-\frac{\D-J}{2}\right)_\b
\nn
&\times{}_4F_3\!\left(
\begin{matrix}
-J,\ \nu,\ 1-\D_\f+\frac{\D-J}{2},\
\D_\f-\frac{\D+J}{2}+\a\\
1-\nu-J,\ 1-\D_\f+\frac{\D-J}{2}-\b,\
\D_\f-\frac{\D+J}{2}
\end{matrix};1\right),
\label{monomial kinematic kernel}
\end{align}
This gives \eqref{unphysical coefficient} and
\eqref{unphysical coefficient fermion}, in agreement with
Ref.~\cite{Barrat:2026jfg}.  Only even $\a+\b$ survives image pairing.
At exceptional parameters, the entire hypergeometric product is continued;
the finite sum \eqref{monomial finite sum kernel} avoids artificial $0/0$
singularities.

\subsection{Projection onto double-twist blocks}
\label{app:double-twist-projection}

Write $z=\rho e^{i\theta}$, $\zb=\rho e^{-i\theta}$,
$\eta=\cos\theta$.  At total degree $L$, a symmetric monomial pair is
$2\rho^L T_m(\eta)$, $m=|\a-\b|>0$; the central monomial is $\rho^L$.
The Chebyshev--Gegenbauer connection formula is
\begin{align}
 2T_m(\eta)=\sum_{r=0}^{\lfloor m/2\rfloor}
 \frac{m\Gamma(m-r)(-\nu)_r(m-2r+\nu)}
 {r!(\nu)_{m-r+1}}\,C_{m-2r}^{(\nu)}(\eta).
 \label{Chebyshev Gegenbauer projection}
\end{align}
Setting $m=\widehat J+2r$ in this finite identity gives the outer
coefficient in \eqref{double twist kinematic kernel}.  The central
monomial supplies its $(\widehat J,r)=(0,0)$ limit.
Conversely, \eqref{double twist homogeneous polynomial} gives
\eqref{basis change matrix}, hence
\begin{align}
H_{\D,J}^{\a,L-\a}
=\sum_{\substack{\widehat n\geq0,\ \widehat J\in2\bb Z_{\geq0}\\
                  2\widehat n+\widehat J=L}}
M_{\widehat n,\widehat J}^{\a,L-\a}
\mathcal Q_{\D,J}^{\widehat n,\widehat J},
\qquad L\in2\bb Z_{\geq0}.
\label{kinematic kernel basis identity}
\end{align}
The sources obey the same transformation:
\begin{align}
B^{\a,\b}
&=\sum_{\widehat n,\widehat J}
M_{\widehat n,\widehat J}^{\a,\b}
\mathcal B^{\widehat n,\widehat J},
&
B_{\rm F}^{\a,\b}
&=\sum_{\widehat n,\widehat J}
M_{\widehat n,\widehat J}^{\a,\b}
\mathcal B_{\rm F}^{\widehat n,\widehat J}.
\label{source basis change}
\end{align}
At each even $L$, all sums are finite and restricted to
$2\widehat n+\widehat J=\a+\b=L$.  The triangular matrix $M$ is invertible
on the symmetric polynomial subspace, establishing the equivalence of
the two bootstrap hierarchies.

\subsection{Fixed-level large-spin expansion}
\label{app:fixed-n-large-spin}

For fixed input $(\Delta,\ell)$, let
\begin{align*}
 p&=\Delta_\phi-\frac{\Delta-\ell}{2},&
 q_k&=p-k,&t_k&=p-\ell+k,&
 w_k&=\frac{(\nu)_k(\nu)_{\ell-k}}{k!(\ell-k)!}.
\end{align*}
Reversing $k\mapsto\ell-k$ in \eqref{double twist kinematic kernel} and
using the Pochhammer form of each binomial gives, for positive even $J$,
\begin{align}
 \mathcal Q_{\Delta,\ell}^{n,J}
 &=\sum_{k=0}^{\ell}\sum_{r=0}^{n}
 w_k\frac{(q_k)_J}{(\nu)_J}
 \frac{(-\nu)_r(t_k)_{n-r}}{r!(n-r)!}\,F_{n,r}(J;q_k),
 \label{fixed n exact Pochhammer kernel}\\
 F_{n,r}(J;q)
 &=\frac{(J+2r)(J+q)_{n+r}}
 {(J+r)_{n+1}(J+\nu+1)_r}
 =1+\frac{n(q-1)+r(q-\nu)}{J}+O(J^{-2}).
 \label{fixed n rational factor}
\end{align}
At fixed $n$, the finite sums obey the Vandermonde identities
\begin{align}
 \sum_{r=0}^{n}\frac{(-\nu)_r(t)_{n-r}}{r!(n-r)!}
 &=\frac{(t-\nu)_n}{n!},\nn
 \sum_{r=0}^{n}\frac{r(-\nu)_r(t)_{n-r}}{r!(n-r)!}
 &=-\nu\frac{(t-\nu+1)_{n-1}}{(n-1)!}\qquad(n\geq1),
 \label{fixed n Vandermonde identities}
\end{align}
with the second sum zero at $n=0$.  In the $k$-sum, only $k=0,1$
contribute through relative order $1/J$.  Combining
\eqref{large spin gamma ratio expansion} with
\eqref{fixed n rational factor}--\eqref{fixed n Vandermonde identities}
gives \eqref{large spin fixed level asymptotic} and the expressions for
$h_n,b_n$ stated below it.

For scalar input, the same result follows from the generating function
and Gegenbauer connection formula
\begin{align}
 (1-2t\eta+t^2)^{-p}&=\sum_{L\geq0}C_L^{(p)}(\eta)t^L,\nn
 C_L^{(p)}(\eta)
 &=\sum_{n=0}^{\lfloor L/2\rfloor}
 \frac{(p)_{L-n}(p-\nu)_n(L-2n+\nu)}
 {n!(\nu)_{L-n+1}}\,C_{L-2n}^{(\nu)}(\eta).
 \label{scalar Gegenbauer connection}
\end{align}
Taking $L=J+2n$, the coefficient is
\begin{align*}
 \frac{(p)_{J+n}(p-\nu)_n(J+\nu)}
 {n!(\nu)_{J+n+1}}
 =\frac{(p)_J}{(\nu)_J}
 \frac{(p-\nu)_n}{n!}
 \frac{(J+p)_n}{(J+\nu+1)_n},
\end{align*}
which proves \eqref{large spin scalar exchange exact} after image summation.
For terminating parameters or vanishing leading coefficients, use these
exact expressions before taking the large-$J$ limit.

\section{Detailed zero-separation benchmarks}
\label{app:zero-distance}

This appendix records the computations behind two complementary tests of
section~\ref{sec:zero-distance}.  The first has a convergent reconstruction in a
specified subtraction scheme.  The second illustrates analytic continuation of a
divergent spectral sum and the distinction between a conformal-vacuum cylinder
correlator and a thermal correlator.

\subsection{The critical large-\texorpdfstring{$N$}{N} \texorpdfstring{$O(N)$}{O(N)} model}
\label{app:ON-zero-distance}

In the normalization of section~\ref{sec:ON-largeN-block-example}, the
three-dimensional propagator at zero spatial separation at leading order in $1/N$ is
\begin{align}
 g_m(\tau)=\sum_{r\in\mathbb Z}
 \frac{e^{-m|\tau+r|}}{|\tau+r|},\qquad 0<\tau<1,\qquad \D_\f=\frac12.
 \label{ON diagonal propagator}
\end{align}
We first leave the dimensionless mass $m>0$ arbitrary.  Criticality subsequently
fixes it to the thermal saddle
\cite{Sachdev:1992py,Chubukov:1993critical,Petkou:2018ynm}.
Expanding the zero image and pairing the other images gives $a_0=1$ and, for
integer $\D\geq1$,
\begin{align}
 a_\D&=\frac{(-m)^\D}{\Gamma(\D+1)}
 +\big[1+(-1)^{\D-1}\big]
 \sum_{k=0}^{\D-1}\frac{m^k}{k!}
 \operatorname{Li}_{\D-k}(e^{-m}).
 \label{ON corrected diagonal coefficients}
\end{align}
In particular, $a_{2p}=m^{2p}/(2p)!$.  The dimension-one singlet is absent in
the critical theory; its vanishing coefficient gives the gap equation
\begin{align}
 a_1=-m+2\operatorname{Li}_1(e^{-m})=0,\qquad
 m=\mth=2\log\frac{1+\sqrt5}{2}.
 \label{ON diagonal gap}
\end{align}
The appearance of the golden ratio and associated thermal polylogarithms is familiar
from the large-$N$ solution \cite{Sachdev:1992py,Sachdev:1993polylog}.
The zero-separation coefficients combine all spins at fixed dimension and do not replace
the spin-resolved data in section~\ref{sec:ON-largeN-block-example}.

An all-level check is possible without truncating the bootstrap equations.
At $\D_\f=1/2$, the identity block has a pole at $s=1$; choose the finite part
that removes $2/(s-1)$ from the sum of Hurwitz zeta functions.  Then
\begin{align}
 (P_0^{\rm B})_{\rm FP}(\tau)&=-\psi(\tau)-\psi(1-\tau),\nn
 P_{2p}^{\rm B}(\tau)&=-\frac{B_{2p}(\tau)}{p}\quad(p\geq1),\qquad
 P_{2j+1}^{\rm B}(\tau)=0\quad(j\geq0),
 \label{ON Bernoulli blocks}
\end{align}
where $\psi$ is the digamma function and $B_n(\tau)$ are Bernoulli polynomials.
Thus the regularized spectral sum is
\begin{align}
 H_m(\tau)\equiv\sum_\D a_\D P_\D^{\rm B}(\tau)
 =-\psi(\tau)-\psi(1-\tau)
 -\sum_{p=1}^{\infty}\frac{m^{2p}B_{2p}(\tau)}{p(2p)!}.
 \label{ON diagonal reconstruction}
\end{align}
This last series converges for $|m|<2\pi$, which includes the physical thermal mass.
The even part of the Bernoulli generating function yields
\begin{align}
 \partial_m H_m(\tau)&=\frac2m-
 \frac{e^{m\tau}+e^{m(1-\tau)}}{e^m-1},\qquad
 \partial_m g_m(\tau)=-
 \frac{e^{m\tau}+e^{m(1-\tau)}}{e^m-1}.
 \label{ON diagonal mass derivatives}
\end{align}
Using
$g_m(\tau)=-2\log m-\psi(\tau)-\psi(1-\tau)-2\gamma_E+o(1)$ as
$m\to0^+$ fixes the integration constant:
\begin{align}
 g_m(\tau)=H_m(\tau)+\kappa,\qquad
 \kappa=-2\log m-2\gamma_E.
 \label{ON diagonal subtraction}
\end{align}
This proves every positive-even-level cancellation and the level-zero equation in
the stated finite-part prescription.  It also separates the subtraction $\kappa$
from the direct-OPE contribution $-m$ in \eqref{ON diagonal gap}.  The latter is
canceled by the nonzero images at the critical mass and is not the dispersive
subtraction.

An equivalent useful form displays how the analytic-sector coefficients are fixed:
\begin{align}
 a_{2\ell+1}
 =\sum_{p=0}^{\infty}\frac{m^{2p}}{(2p)!}
 (A_{2p}^{2\ell})_{\rm B},\qquad \ell\geq1,
 \qquad (A_{2j+1}^{2\ell})_{\rm B}=-\delta_{j,\ell}.
 \label{ON diagonal triangular solution}
\end{align}
The potentially misleading term at $p=\ell$ must be evaluated as a product:
\begin{align}
 (A_{2\ell}^{2\ell})_{\rm B}
 =\lim_{\epsilon\to0}
 \frac{2(1-2\ell+\epsilon)_{2\ell}}{(2\ell)!}\zeta(1+\epsilon)
 =-\frac1\ell.
 \label{ON diagonal removable pole}
\end{align}
Discarding the Pochhammer zero before taking the zeta-pole limit would spoil the
bootstrap equation.  Finally, \eqref{ON diagonal subtraction} holds for any
$0<m<2\pi$: the image reconstruction alone does not impose criticality.
The physical spectral absence condition \eqref{ON diagonal gap} is indispensable.

\subsection{A resummed Lee--Yang cylinder identity}
\label{app:LY-zero-distance}

The nonunitary minimal model $M(2,5)$ has $c=-22/5$ and a scalar primary of dimension
$\D_\f=-2/5$ \cite{Belavin:1984cft,Cardy:1985leeYang}.
The conformal map from the plane gives the conformal-vacuum cylinder correlator
\begin{align}
 g_{\rm{vac}}(\tau)=\left(\frac{\sin\pi\tau}{\pi}\right)^\delta,
 \qquad \delta=\frac45,\qquad 0<\tau<1.
 \label{LY 2pt x=0}
\end{align}
In a nonunitary CFT the identity state need not be the lowest-energy cylinder
state \cite{Gannon:2003nonunitary}.  Here the state created by $\phi$ has lower
energy.  Exchanging the Euclidean cylinder channels to obtain the infinite-line
thermal correlator therefore selects that state, not the conformal vacuum.
The two cylinder correlators are explicitly distinguished in
Ref.~\cite{ArguelloCruz:2026yanglee}.
We use \eqref{LY 2pt x=0} only as an analytic cancellation test, as motivated by
the cylinder example of Ref.~\cite{Barrat:2025nvu}; it is not a verification for
the physical Lee--Yang Gibbs state.

To find the OPE coefficients, write
\begin{align}
 \tau^{-\delta}g_{\rm{vac}}(\tau)
 =\left(\frac{\sin\pi\tau}{\pi\tau}\right)^\delta
 =\sum_{n=0}^{\infty}a_{2n}\tau^{2n},\qquad a_0=1.
 \label{LY coefficient generating function}
\end{align}
Taking a logarithm of the sine product gives
$\log(\tau^{-\delta}g_{\rm{vac}})=-\delta\sum_{k\geq1}\zeta(2k)\tau^{2k}/k$.
Differentiating and matching powers hence yields
\begin{align}
 a_{2n}=-\frac{\delta}{n}\sum_{k=1}^n\zeta(2k)a_{2n-2k},
 \qquad a_2=-\frac{2\pi^2}{15}.
 \label{LY corrected recursion}
\end{align}
Equivalently, $a_\D=-8(5\D)^{-1}
\sum_{k=1}^{\D/2}\zeta(2k)a_{\D-2k}$ for positive even $\D$.

The ordinary spectral sums of the kernels are factorially divergent.
KMS reflection of \eqref{LY 2pt x=0} is therefore not sufficient to justify
term-by-term cancellation.  An explicit common prescription makes the statement
precise.  For even $L$, the zeta functional equation gives
\begin{align}
 \operatorname{Reg}\sum_{n=0}^{\infty}a_{2n}(A_{2n}^L)_{\rm B}
 &=c_L S_L,\qquad
 c_L=\frac{(-1)^{L/2+1}}{L!}
 \frac{2\sin(\pi\delta/2)}{\pi}(2\pi)^{L-\delta},\nn
 S_L&=\operatorname{Reg}\sum_{n=0}^{\infty}
 \frac{(-1)^na_{2n}}{(2\pi)^{2n}}
 \Gamma(2n+1+\delta)\zeta(2n+1+\delta-L).
 \label{LY reflected sum}
\end{align}
A Borel--Leroy reconstruction removes the displayed gamma-factor growth.
Reconstructing the coefficient generating function on the imaginary axis and
summing the zeta images leads to the prescription
\begin{align}
 S_L\equiv\operatorname{FP}\int_0^\infty
 \big[2\sinh(t/2)\big]^\delta\operatorname{Li}_{-L}(e^{-t})\,{\rm d}t.
 \label{LY finite part integral}
\end{align}
Here FP is defined by inserting $t^\epsilon$, evaluating for
$\operatorname{Re}\epsilon>L-\delta$, and analytically continuing to
$\epsilon=0$.  The integral converges at $t\to\infty$ since $\delta<2$.
The noninteger powers near $t=0$ at $\delta=4/5$ make this continuation regular at
$\epsilon=0$.  This supplies a definite prescription even when the unregulated
integral and the original spectral sum do not converge.

This prescription also has a convergent subtracted form.  Set
$b_n=(-1)^na_{2n}/(2\pi)^{2n}$ and
$F(t)=[2\sinh(t/2)/t]^\delta=\sum_{n\geq0}b_nt^{2n}$ near the origin.
For $L=2q\geq2$,
\begin{align*}
 S_{2q}={}&\sum_{n=0}^{q-1}b_n\Gamma(2n+1+\delta)
             \zeta(2n+1+\delta-2q)\\
 &+\int_0^\infty t^\delta
 \left[F(t)-\sum_{n=0}^{q-1}b_nt^{2n}\right]
 \operatorname{Li}_{-2q}(e^{-t})\,\mathrm dt .
\end{align*}
The integrand is $O(t^{\delta-1})$ at zero, and the integral converges
exponentially at infinity
for $0<\delta<2$.  Mellin continuation of each subtracted monomial gives the
displayed $\Gamma\zeta$ term.  This is a fixed-level resummation, not a claim of
uniform convergence of the full spectral reconstruction.

There is an exact evaluation.  For $L\geq1$, use the Eulerian polynomial,
\begin{align}
 \operatorname{Li}_{-L}(x)
 =\frac{x}{(1-x)^{L+1}}
 \sum_{j=0}^{L-1}\genfrac{\langle}{\rangle}{0pt}{}{L}{j} x^j ,
 \label{LY Eulerian polylogarithm}
\end{align}
and set $x=e^{-t}$ in \eqref{LY finite part integral}.  To compare the local subtraction
prescriptions near $x=1$, write $(-\log x)^\epsilon=(1-x)^\epsilon h(x)^\epsilon$,
where $h(x)=-\log x/(1-x)$ is analytic near $x=1$ and $h(1)=1$.
The continued integral has no regulator pole at $\epsilon=0$ for $\delta=4/5$, so the
$O(\epsilon)$ change of regulator cannot change the continued value.
Continuing the power of $(1-x)$ at fixed $\delta$ therefore gives
\begin{align}
 S_L=\sum_{j=0}^{L-1}
 \genfrac{\langle}{\rangle}{0pt}{}{L}{j}
 B\left(j+1-\frac{\delta}{2},\,\delta-L\right).
 \label{LY beta sum}
\end{align}
For even $L$, the Eulerian numbers are invariant under $j\mapsto L-1-j$,
whereas gamma reflection gives
\begin{align}
 B\left(j+1-\frac{\delta}{2},\delta-L\right)
 =-B\left(L-j-\frac{\delta}{2},\delta-L\right).
 \label{LY beta reflection}
\end{align}
Every pair cancels, proving $S_L=0$ for $L=2,4,\ldots$.
At level zero the integral converges and instead gives
$S_0=B(1-\delta/2,\delta)$, so that the sourced equation fixes
$\kappa=-c_0 B(1-\delta/2,\delta)$ in this prescription.
The example thus verifies the claimed positive-level identities, while making
explicit the resummation and the state choice on which the statement depends.

\bibliographystyle{JHEP}
\bibliography{references}

\end{document}